\documentstyle[12pt,epsfig]{article}
\renewcommand{\thesubsection}{\arabic{subsection}}

\renewcommand{\theequation}{\arabic{subsection}.\arabic{equation}}
\renewcommand{\thefootnote}{*}
\begin{document}
\hspace*{11.5cm} KA-THEP-4-1995

\vspace*{2.0cm}
\begin{center} \Large
 Fermionic decays of neutral MSSM Higgs bosons \\
 at the one-loop level\footnote{
 Supported in part by the European Union under contract CHRX-CT92-0004.}
\renewcommand{\thefootnote}{\arabic{footnote}}
\setcounter{footnote}{0}
\\[1.0cm]
 \large
 A. Dabelstein\footnote{
 E-Mail: ADD@DMUMPIWH.BITNET} \\
 \vspace{0.8cm}
 \normalsize \sl
 Institut f\"ur Theoretische Physik\\
 Universit\"at Karlsruhe\\
 Kaiserstr. 12\\
 D-76128 Karlsruhe, Germany \\
\end{center}
\rm
\vspace{0.3cm}
\begin{abstract}
\vspace*{0.1cm} \hspace{0.5cm}
The results of a complete one-loop calculation for the fermionic
decay width $\Gamma (h^0, H^0, A^0 \rightarrow f \bar{f})$ of the
neutral MSSM Higgs bosons are presented and the dominant light Higgs
decay channel $h^0 \rightarrow b \bar{b}$ is discussed in detail.
The enhancement of
$\Gamma (h^0 \rightarrow b \bar{b})$ compared to the standard Higgs
decay is shown for pseudoscalar masses $M_A \le 300$ GeV, where
the one-loop contributions in the MSSM and SM are different.
Simpler approximation formulae for the Higgs decays are given and their
quality is discussed by introducing an effective neutral scalar mixing angle
$\sin^2 \alpha_{eff}$. Finally the Higgs branching ratios in $b \bar{b}$,
$c \bar{c}$, $\tau^+ \tau^-$ are calculated.
\end{abstract}
\normalsize
\renewcommand{\thepage}{}
\newpage
\setcounter{page}{1}
%
%
\vspace*{0.1cm} \hspace*{0.5cm}
\renewcommand{\thepage}{\arabic{page}}
%
%
\subsection{Introduction}
\setcounter{equation}{0}\setcounter{footnote}{0}
\vspace*{0.1cm} \hspace*{0.5cm}
Supersymmetry is at present the most predictive framework for physics
beyond the standard model \cite{hunter}. One theoretical motivation is the
cancellation of quadratically divergent contributions to the mass of the
scalar Higgs particle. This problem of naturalness is solved in
supersymmetric theories. Supersymmetric models allow the unification
of gauge couplings at the GUT scale $\cal{O} \rm (10^{15}$ GeV $)$
\cite{boer}.
\par \medskip
The minimal supersymmetric standard model (MSSM) is considered as the most
general supersymmetric extension of the standard model (SM) at low energies
\cite{hara}.
The Higgs sector of the MSSM is that of a 2-Higgs-doublet model, where the
coefficients of the Higgs potential are restricted by supersymmetry.
As a consequence of the supersymmetric Higgs potential, a light Higgs boson
exists with a tree level upper mass bound given by the $Z^0$ mass.
Radiative corrections to the Higgs mass spectrum, however, predict
an upper limit of the light Higgs mass $\cal{O}\rm $($130$ GeV)
\cite{hempf1, upper1}.
Calculations were performed at the one-loop level using renormalization
group technique \cite{barb1}, effective potential approximation \cite{ellis}
and one-loop calculations with top and stop contributions \cite{okada,yama}.
Two-loop effects to the upper limit of the lightest Higgs boson
mass are discussed in \cite{upper2}.
\par \medskip
Production and decay properties of the Higgs boson are charcteristic
quantities for the experimental Higgs search at LEP 2, at a $500$ GeV
$e^+ e^-$ collider  and LHC. Precise
predictions of these quantities require the inclusion of radiative
corrections. The one-loop Higgs production cross section and their
respective branching ratios (decay width) at $e^+ e^-$ and
$p p$ colliders may allow to distinguish between
a standard or MSSM Higgs sector. As a first step, complete on-shell
renormalization schemes for the MSSM Higgs sector were presented
\cite{okada,yama,pok1,dabx}.
\par \medskip
In this article the complete one-loop partial decay width of the neutral
MSSM Higgs bosons $h^0$ ($H^0, A^0$) $\rightarrow f \bar{f}$ is calculated
within the on-shell scheme \cite{dabx}.
The dominant fermionic decay width of the light MSSM
Higgs boson is discussed in detail and the branching ratios of the light
scalar Higgs $h^0$ in $b \bar{b}$, $c \bar{c}$, $\tau^+ \tau^-$ are
presented. The discussion points out the differences of the MSSM and the
standard Higgs decay width and branching ratios.
\par \medskip
Section 2 presents an overview of the Higgs production and decay mechanisms,
including radiative corrections. The fermionic decay width with full
one-loop corrections is calculated in section 3. Finally the numerical results
for the decay width and branching ratios are discussed in section 4 for
QED/QCD, gluino and weak MSSM virtual contributions.
Vertex corrections and self energies are given in the appendix.
\subsection{Decay channels of the neutral Higgs boson}
\setcounter{equation}{0}\setcounter{footnote}{0}
\vspace*{0.1cm} \hspace*{0.5cm}
Once a Higgs boson is found, it is of importance to investigate its
characteristic decay properties.
The observed cross section is composed by the production cross section
for $e^+ e^- \rightarrow Z h (H)$ or $ e^+ e^- \rightarrow A h (H)$
and the branching ratios for the subsequent decays of the scalar bosons.
The decay width (respectively the branching ratios) as well as the
mass-width correlation are the quantities to differentiate between
Higgs bosons of various origin. In the following we briefly review
the decay modes of the neutral MSSM Higgs particles and discuss the
most important fermionic decays in some more detail.
Except from a small part of the parameter space, the fermionic
decays are the only decay modes of the light Higgs allowed at tree
level. The bosonic decays $ H^0 \rightarrow Z Z, W W$ (which are
dominant above the threshold in case of the standard Higgs)
are kept small also for increasing $M_{H^0}$ by the factor $ \cos^2 (\alpha
- \beta )$.
\smallskip
\par
In Tab. 1 we list the various decay channels of the neutral Higgs
bosons indicating the level of the theoretical predictions by:
\medskip
\par
\begin{tabular}[h]{rl}
full electroweak: &
 complete 1-loop electroweak calculation performed and available\\
QCD: & QCD corrections performed and available \\
improved Born: &
decay width is calculated including the complete 1-loop scalar
2-\\ & point functions. \\
\end{tabular}
The signature $\circ$ denotes the corresponding decay mode as
proceeding through 1-loop in lowest order.
%
%
\scriptsize
\begin{table}
\begin{flushleft}
\caption{Decay channels of the neutral Higgs bosons}
\begin{tabular}[t]{|l|l|l|l|l|l|l|l|}
\hline
& $f \bar{f}$ & $ Z Z$ & $ Z \gamma $ & $ \gamma \gamma $ &
$ h h $ & $ A A $ & $ Z h $ \\
\hline
$h^0 \rightarrow$ & full EW$^{1)}$ & full EW$^{2)}$ & $\circ$
 full EW$^{2)}$ & $\circ$ full EW$^{2)}$ & - & full EW$^{2)}$ & -  \\
& QCD$^{3)}$ & & QCD$^{4)}$ & QCD$^{5)}$ & & &  \\
\hline
$H^0 \rightarrow$ & full EW$^{1)}$ & full EW$^{2)}$ & $\circ$
 full EW$^{2)}$ & $\circ$ full EW$^{2)}$ & full EW$^{2)}$ &
 full EW$^{2)}$ & - \\
& QCD$^{3)}$ & & QCD$^{4)}$ & QCD$^{5)}$ & & &  \\
\hline
$A^0 \rightarrow$ & full EW$^{1)}$ & $\circ$ full EW$^{2)}$ &
$\circ$ full EW$^{2)}$ & $\circ$ full EW$^{2)}$ & - & - &
full EW$^{2)}$  \\
& QCD$^{3)}$ & & & & & & \\
\hline
\end{tabular}
\begin{tabular}[t]{|l|l|l|l|}
\hline
& $\tilde{\chi} \tilde{\chi}$ & $g g$  & $ \tilde{g} \tilde{g}$  \\
\hline
$h^0 \rightarrow$ & improved & $\circ$ full EW$^{2)}$  & $\circ$ QCD$^{7)}$ \\
& Born$^{6)}$ & QCD$^{5)}$ &  \\
\hline
$H^0 \rightarrow$ & improved & $\circ$ full EW$^{2)}$  & $\circ$ QCD$^{7})$ \\
& Born$^{6)}$ & QCD$^{5)}$ &  \\
\hline
$A^0 \rightarrow$ & improved & $\circ$ full EW$^{2)}$ & $\circ$ QCD$^{7)}$ \\
& Born$^{6)}$ & QCD$^{5)}$ &  \\
\hline
\end{tabular}
\end{flushleft}
%
\noindent
$^{1)}$ Dabelstein, Hollik \\
$^{2)}$ Chankowski, Pokorski, Rosiek \cite{pok1,pok6} \\
$^{3)}$ Braaten, Leveille \cite{braaten}; Bardin et al. \cite{bardin};
        Drees, Hikasa \cite{Drees2}, Chankowski et al. \cite{pok6} \\
$^{4)}$ Djouadi, Spira, van der Bij, Zerwas
        \cite{spira1} \\
$^{5)}$ Djouadi, Spira, Zerwas \cite{spira2} \\
$^{6)}$ Gunion, Haber \cite{haber2}, improved by the Higgs 2-point functions
from Chankowski, Pokorski, Rosiek \cite{pok6} \\
$^{7)}$ Ng, Pois, Yuan; Djouadi, Drees \cite{gluinos}
\end{table}
%
%
\subsection{Fermionic Higgs decays}
\setcounter{equation}{0}\setcounter{footnote}{0}
\normalsize
\subsubsection{Tree level structure}
\vspace*{0.1cm} \hspace*{0.5cm}
The Higgs sector of the MSSM consists of two scalar doublets $H_1$, $H_2$
with opposite hypercharge $Y_1 = -Y_2 = -1$ and vacuum expectation
values $v_1, v_2$ \cite{hunter,dabx}.
The Higgs potential contains two independent free
parameters, which can conveniently be chosen as $\tan \beta = v_2 / v_1$
and $M_A$, where $M_A$ is the mass of the $A^0$ boson.
\par \smallskip
In lowest order the Yukawa coupling of the standard (MSSM) Higgs
bosons to fermions reads:
\begin{equation}
T_{Hff} = -\frac{i e m_f}{2 s_W M_W} \cdot \kappa_H^f \ ,
\end{equation}
where the weak mixing angle $s_W = \sin \theta_W$
is introduced in the convention of $s_W^2 = 1 - M_W^2/M_Z^2$ \cite{sirlin}.
The coefficients $\kappa_H^f$ are listed in Tab. \ref{tab1} for
the neutral Higgs particles of the standard model ($H = SM$) and the MSSM
($H = h^0,H^0, A^0$).
\setcounter{table}{1}
\begin{table}[ht]
\begin{center}
 \begin{tabular}[b]{|c|c|c|c|c|}
 \hline
 $\kappa_H^f$ & $H=$ SM-Higgs & $H = h^0$ & $H = H^0$ & $H = A^0$  \\[0.2cm]
 \hline
 $f = u$ & 1 & $\frac{\cos \alpha}{\sin \beta}$ & $ \frac{\sin \alpha}{\sin
\beta}$ & $ - i \cot \beta \ \gamma_5 $ \\[0.2cm]
 \hline
 $f = d$ & 1 & $-\frac{\sin \alpha}{\cos \beta}$ & $\frac{\cos \alpha}{\cos
\beta}$ &  $- i \tan \beta \ \gamma_5$ \\[0.2cm]
 \hline
 \end{tabular}
\end{center}
\caption{Coefficients of the $H \rightarrow f \bar{f}$ vertex}
\label{tab1}
\end{table}
In lowest order, the partial decay width $H \rightarrow f \bar{f}$
of a neutral Higgs boson $H$ can be written  in the following way:
\begin{equation}
\Gamma_0 ( H  \rightarrow f \bar{f} )= \frac{N_C G_F \
 m_f^2} {4 \sqrt{2} \pi } \ \tilde{\beta}^n \ m_H \ | \kappa_H^f |^2 \ ,
\label{glhffgf}
\end{equation}
where $n = 3,3,3,1$ for $ H = H_{SM}, h^0, H^0, A^0$. In the following
$H$ always denotes one of the neutral Higgs particles, and
$$ \tilde{\beta} = \sqrt{1 - \frac{4 m_f^2}{m_H^2} } \ . $$
$G_F$ is the Fermi constant, related to $M_W$ by
\begin{equation}
 G_F = \frac{\pi \alpha}{\sqrt{2} s_W^2 M_W^2} \cdot \frac{1}{1 -
\Delta r} \ ,
\end{equation}
where $\Delta r $ is the (SM or MSSM) radiative correction to the
$\mu^-$ decay amplitude \cite{sola}.
\par \smallskip
Radiative corrections in the MSSM Higgs sector modify the tree-level
decay rate Eq. (\ref{glhffgf}) substantially, with the main effect from
loops involving the top quark and its scalar partner $\tilde{t}$.
The complete decay width comprises the following radiative corrections to
\begin{itemize}
\item[(i)]   the physical neutral scalar MSSM Higgs
\item[(ii)]  the full one-loop decay amplitude $H \rightarrow f \bar{f}$
\item[(iii)] $\Delta r$ in the MSSM.
\end{itemize}
\subsubsection{One-loop structure $H \rightarrow f \bar{f}$}
\vspace*{0.1cm} \hspace*{0.5cm}
In this article the one-loop decay amplitudes $H \rightarrow f \bar{f}$
are calculated within the complete one-loop renormalization scheme
for the MSSM Higgs sector described in \cite{dabx}.
The one-loop decay widths for $h^0, H^0, A^0 \rightarrow f \bar{f}$ with
full electroweak MSSM corrections read: \\
\begin{eqnarray}
 \Gamma_1 (h^0 \rightarrow f \bar{f} ) & = & \frac{N_C \,G_F \,m_f^2 }
 { 4 \sqrt{2} \pi  } M_{h^0} \tilde{\beta}^3 \ |\kappa_{h^0}^f|^2
 \cdot \nonumber \\ & &  Z_{h^0} \  \left( \
 | 1 + Z_{h^0H^0} \frac{ \kappa_{H^0}^f}{\kappa_{h^0}^f} |^2
 + 2 \Re e( 1 + Z_{h^0H^0}
 \frac{ \kappa_{H^0}^f}{\kappa_{h^0}^f}) \Delta T_{h^0} \ \right)
 \ ( 1 - \Delta r_{MSSM} ) \nonumber \\[0.2cm]
 \Gamma_1 (H^0 \rightarrow f \bar{f} ) & = & \frac{N_C \, G_F \, m_f^2 }
 { 4 \sqrt{2} \pi} M_{H^0} \tilde{\beta}^3 \ |\kappa_{H^0}^f|^2 \cdot \nonumber
 \\[0.2cm]  & &  Z_{H^0} \ \left( \
 | 1 +  Z_{H^0h^0} \frac{ \kappa_{h^0}^f}{\kappa_{H^0}^f} |^2
 + 2 \Re e( 1 + Z_{H^0h^0}
 \frac{ \kappa_{h^0}^f}{ \kappa_{H^0}^f} ) \Delta T_{H^0} \ \right)
 \ ( 1 - \Delta r_{MSSM} ) \nonumber \\[0.2cm]
 \Gamma_1 (A^0 \rightarrow f \bar{f} ) & = & \frac{N_C \, G_F \, m_f^2 }
 { 4 \sqrt{2} \pi } M_{A^0} \tilde{\beta} \ |\kappa_{A^0}^f|^2 \ \left(
 \ 1 + 2 \Re e \ \Delta  T_{A^0}  \ \right) \ ( 1 - \Delta r_{MSSM} ) \ .
 \nonumber \\
\label{gamma1}
\end{eqnarray}
$Z_{h^0}$ and $Z_{H^0}$ are finite wave function renormalizations
of the external light and heavy Higgs particles:
\begin{eqnarray}
 Z_{h^0} & = & Res_{M_{h^0}} \ \Delta_{h^0}  = \frac{1}{ 1
+ \hat{  \Sigma}_{h^0}'
 (k^2)  - \left( \frac{  \hat{ \Sigma}_{h^0H^0}^2 (k^2) }{k^2 - m_{H^0}^2 +
 \hat{ \Sigma}_{H^0} (k^2) }  \right ) ' } \ |
 _{k^2 =    M_{h^0}^2}  \nonumber \\
 Z_{H^0} & = & Res_{M_{H^0}} \ \Delta_{H^0}  =  \frac{1}{ 1
 + \hat{  \Sigma}_{H^0}'
 (k^2)  - \left( \frac{  \hat{ \Sigma}_{h^0H^0}^2 (k^2) }{k^2 - m_{h^0}^2 +
 \hat{ \Sigma}_{h^0} (k^2) }  \right ) ' } \ |
 _{k^2 =    M_{H^0}^2}  \nonumber   \ ,
\end{eqnarray}
where $\Delta_{h^0}$, $\Delta_{H^0}$ are the diagonal $h^0$, $H^0$-propagators
and $\hat{\Sigma}_{h^0}$, $\hat{\Sigma}_{H^0}$, $\hat{\Sigma}_{hH}$ are
the renormalized $h^0$, $H^0$ self energies and mixing \cite{dabx}.
The $h^0$-$H^0$ mixing enters in terms of
\begin{eqnarray}
 Z_{h^0H^0} & = & - \frac{ \hat{\Sigma}_{h^0H^0}
 (M^2_{h^0})}{M^2_{h^0} - m_{H^0}^2 + \hat{\Sigma}_{H^0} (M^2_{h^0}) }
  \nonumber \\
 Z_{H^0h^0} & = & - \frac{ \hat{\Sigma}_{h^0H^0}
 (M^2_{H^0})}{M^2_{H^0} - m_{h^0}^2 + \hat{\Sigma}_{h^0} (M^2_{H^0}) } \ .
 \nonumber
\end{eqnarray}
The pseudoscalar Higgs self energy and $A^0 G^0$ mixing in the decay width
$\Gamma_1 ( A^0 \rightarrow f \bar{f} )$,
Eq. (\ref{gamma1}), do not contribute, because the renormalization condition
sets the residue of the pseudoscalar Higgs propagator equal to one.
The Slavnov-Taylor identity yields:
\begin{equation}
 k^2 \hat{\Sigma}^{A^0Z^0} (k^2) - M_Z \hat{\Sigma}^{A^0G^0} (k^2) = 0 \ ,
\label{glstaz}
\end{equation}
where $\hat{\Sigma}^{A^0Z^0} (M_A^2) = 0$ \cite{dabx}.
\par \smallskip
The renormalized vertex correction $\Delta T_H$ in Eq. (\ref{gamma1})
is the sum of the one-loop vertex diagrams $\Delta T_i$ given in appendix A
and a counterterm CT:
\begin{equation}
 \Delta T_{h^0,H^0,A^0} =  \sum_{i = 1}^{N} \
 ( \ \frac{ \alpha}{4 \pi} \Delta T_i  \ )_{h^0,H^0,A^0}  +  CT \ .
\label{gldelt}
\end{equation}
The counterterm CT reads:
\begin{eqnarray}
CT = \frac{\delta m_f}{m_f} + \frac{\delta Z_L^f + \delta Z_R^f}{2}
 + \frac{\delta v}{v} \ ,
\end{eqnarray}
with the non-universal contribution from the fermion self energies (App. A):
\begin{equation}
 \frac{\delta m_f}
 {m_f}  + \frac{\delta Z_L^f +
 \delta Z_R^f}{2} \ =
 \Sigma_S^f(m_f^2)-2m_f^2 (
\Sigma_S^{'f} (m_f^2) + \Sigma_V^{'f} (m_f^2) \ ) \ ,
\end{equation}
and the universal part:
\begin{eqnarray}
 2 \frac{ \delta v}{v} & = & 2 \frac{ \delta v_i}{v_i} =
 - \Sigma'_A (M_A^2) + \frac{ \tan \beta - \cot \beta }{M_Z }
 \Sigma_{A Z} (M_A^2) - \\ \nonumber & & ( - \Sigma_\gamma' ( 0 ) + 2
 \frac{s_W}{c_W}  \frac{ \Sigma_{ \gamma Z} ( 0 ) }{ M_Z^2 } +
 \frac{c^2_W}{s^2_W}
 \frac{ \Sigma_Z ( M_Z^2 ) }{M_Z^2} - \frac{ c^2_W - s^2_W }{s^2_W }
 \frac{ \Sigma_W ( M_W^2 ) }{M_W^2} \ ) \ .
 \end{eqnarray}
\subsubsection{The one-loop mixing angle $\sin \alpha$}
\vspace*{0.1cm} \hspace*{0.5cm}
Introducing universal one-loop coupling coefficients via
\begin{eqnarray}
 \kappa_{h^0,1-loop}^f & = & \sqrt{ Z_{h^0}} \, ( \kappa_{h^0}^f
    + Z_{h^0H^0} \, \kappa_{H^0}^f ) \nonumber \\
 \kappa_{H^0,1-loop}^f & = & \sqrt{ Z_{H^0}} \, ( \kappa_{H^0}^f
    + Z_{H^0h^0} \, \kappa_{h^0}^f ) \ ,
\label{glkapp}
\end{eqnarray}
\smallskip
an effective universal one-loop mixing angle $\alpha_{1-loop}$ can be defined
through the coefficients in Tab. \ref{tab1} :
\begin{equation}
 \sin^2 \alpha_{1-loop} = \cos^2 \beta \cdot ( \kappa_{h^0,1-loop}^f )^2 \ ,
 \ f = b \ \mbox{and} \ - \pi /2 \le \alpha_{1-loop} \le 0 \ .
\label{glsina1}
\end{equation}
\smallskip \par
A good approximation for this mixing angle $\alpha_{1-loop}$,
including only the leading one-loop term
\begin{eqnarray}
 \omega_t = \frac{N_C G_F m_{t}^4 }{\sqrt{2} \pi^2 \sin^2 \beta} \
  & & \hspace{-0.5cm}
 \left(  \log \ ( \frac{m_{\tilde{t}_1} m_{\tilde{t}_2} }{m_{t}^2} )
 + \frac{A_t ( A_t + \mu \cot \beta)}{ m_{\tilde{t}_1}^2 -
 m_{\tilde{t}_2}^2 } \log \frac{ m_{\tilde{t}_1}^2 }{ m_{\tilde{t}_2}^2 }
 \right. \nonumber \\ & &  \hspace*{-0.5cm}    \left. +
 \frac{A_t^2 ( A_t + \mu \cot \beta)^2 }{ ( m_{\tilde{t}_1}^2 -
 m_{\tilde{t}_2}^2 )^2} \left( 1 - \frac{m_{\tilde{t}_1}^2 + m_{\tilde{t}_2}^2}
 {m_{\tilde{t}_1}^2 - m_{\tilde{t}_2}^2} \log \frac{m_{\tilde{t}_1}}
 {m_{\tilde{t}_2}} \right) \ \right)   \nonumber \\
 \lambda_t = \frac{N_C G_F m_{t}^4 }{\sqrt{2} \pi^2 \sin^2 \beta} \
  & & \hspace{-0.5cm}
 \left( \frac{ \mu ( A_t + \mu \cot \beta)}{ m_{\tilde{t}_1}^2 -
 m_{\tilde{t}_2}^2 } \log \frac{ m_{\tilde{t}_1}^2 }{ m_{\tilde{t}_2}^2 }
 \right. \nonumber \\ & &  \hspace*{-0.5cm}    \left. +
 \frac{2 \mu A_t ( A_t + \mu \cot \beta)^2 }{ ( m_{\tilde{t}_1}^2 -
 m_{\tilde{t}_2}^2 )^2} \left( 1 - \frac{m_{\tilde{t}_1}^2 + m_{\tilde{t}_2}^2}
 {m_{\tilde{t}_1}^2 - m_{\tilde{t}_2}^2} \log \frac{m_{\tilde{t}_1}}
 {m_{\tilde{t}_2}} \right) \ \right)   \nonumber \\
 \sigma_t = \frac{N_C G_F m_{t}^4 }{\sqrt{2} \pi^2 \sin^2 \beta} \
  & & \hspace{-0.5cm}
 \frac{ \mu^2 ( A_t + \mu \cot \beta)^2 }{ ( m_{\tilde{t}_1}^2 -
 m_{\tilde{t}_2}^2 )^2} \left( 1 - \frac{m_{\tilde{t}_1}^2 + m_{\tilde{t}_2}^2}
 {m_{\tilde{t}_1}^2 - m_{\tilde{t}_2}^2} \log \frac{m_{\tilde{t}_1}}
 {m_{\tilde{t}_2}} \right)  \ ,
\label{leadom2}
\end{eqnarray}
follows from the diagonalization of the one-loop Higgs mass matrix
\cite{kunszt}
\begin{equation}
 \cal{M}_{\rm Higgs} \rm = \frac{ \sin 2 \beta }{2}\left( \begin{array}{ll}
     \cot \beta \ M_Z^2 + \tan \beta \ M_A^2 + \sigma_t & - M_Z^2 - M_A^2
     + \lambda_t  \\
     - M_Z^2 - M_A^2 + \lambda_t & \tan \beta \ M_Z^2 + \cot \beta \ M_A^2
     + \omega_t
     \end{array} \right) \ .
\label{glalphaap}
\end{equation}
This approximate effective mixing angle $\alpha_{eff}$ is determined by
\begin{equation}
 \tan \alpha_{eff} = \frac{ - (M_A^2 + M_Z^2 - \lambda_t) \ \tan \beta }{
 M_Z^2 + M_A^2 \tan^2 \beta + \sigma_t \tan \beta - ( 1 + \tan^2 \beta )
 \ M_{h^0,eff}^2 } \ ,
\label{glalpheffx}
\end{equation}
where $M_{h^0,eff}$ is the solution for the light Higgs mass \cite{ellis} :
\begin{eqnarray}
 M^2_{H,h,\, eff} & = & \frac{M_A^2 + M_Z^2 + \omega_t + \sigma_t}{2}
 \pm \, \left( \
 \frac{ (M_A^2 + M_Z^2)^2 + ( \omega_t - \sigma_t )^2}{4}
 - M_A^2 M_Z^2 \cos^2 2\beta \right. \nonumber \\ & & \left. + \
 \frac{(\omega_t - \sigma_t) \cos 2\beta}{2}  (M_A^2 - M_Z^2)
 - \frac{\lambda_t \sin 2\beta}{2} (M_A^2 + M_Z^2) + \frac{\lambda_t^2}{4} \
 \right)^{1/2} \ .
\label{glmapp2} \nonumber \\
\end{eqnarray}
The mixing angle $\alpha_{eff}$ corresponds to the effective potential
approach with $top$ and $stop$ contributions.
\par
Incorporating also the other correction terms of Eq. (\ref{gamma1}),
a complete one-loop effective mixing angle $\alpha_f$ can be defined
\begin{eqnarray}
 \frac{ \Gamma_1 (h^0 \rightarrow f \bar{f}) \, \cos^2 \beta}{\Gamma_{SM, 0}
 (h \rightarrow f \bar{f}) } = \sin^2 \alpha_{1-loop} + \delta_V^f - \Delta
 r_{MSSM} \equiv \sin^2 \alpha_f  & , & ( I_3^f = - 1/2 ) \nonumber \\
 \frac{ \Gamma_1 (h^0 \rightarrow f \bar{f}) \, \sin^2 \beta}{\Gamma_{SM, 0}
 (h \rightarrow f \bar{f}) } = \cos^2 \alpha_{1-loop} + \delta_V^f - \Delta
 r_{MSSM} \equiv \cos^2 \alpha_f \ & , & ( I_3^f = + 1/2 ) \nonumber \\
\label{alphaf}
\end{eqnarray}
with the full one-loop width $\Gamma_1$, Eq. (\ref{gamma1}), normalized
to the standard tree level width $\Gamma_{SM,0}$.
Besides the universal mixing angle $\alpha_1$, defined in
Eq. (\ref{glsina1}), $\Delta r_{MSSM}$ is another universal contribution,
whereas the residual, mainly vertex correction $\delta_V^f$ is fermion
specific making the complete one-loop effective mixing angle flavour
dependent. A momentum dependent mixing angle $\alpha$ has been defined in
\cite{diaz2}.
\par \smallskip
For $f = b$, $\sin^2 \alpha_b$ is shown in Figs. 1 a,b
as a function of the renormalized light Higgs mass $M_{h^0}$ and for fixed
$\tan \beta$ values.
Together with the complete result (dashed line), the approximations in
Eqs. (\ref{glsina1}, \ref{glalpheffx}) are also shown (dotted, full).
In the range $\tan \beta \le 30 $ the full calculation is about
$8 \%$ below the approximation (\ref{glalpheffx}).
Figs. 1 c,d plot $\sin^2 \alpha$ as a function of the pseudoscalar mass
$M_A$ and with the same set of parameters as in Figs. 1 a,b.
The presentation in Figs. 1 c,d shows the complete result and the
approximations  $\sin^2 \alpha_{eff}$, $\sin^2 \alpha_{1-loop}$ for
large pseudoscalar mass $ M_A > 150$ GeV more precisely than Figs. 1 a,b
in the Higgs mass range near the upper limit of the light Higgs mass $M_{h^0}$.
\smallskip
\subsection{Discussion}
\vspace*{0.1cm} \hspace*{0.5cm}
\setcounter{equation}{0}\setcounter{footnote}{0}
Radiative corrections to the partial decay width $H \rightarrow f \bar{f}$
include QED/QCD, weak MSSM and virtual gluino contributions.
The following subsections discuss these corrections and
present the numerical sizes individually. QED/QCD corrections are identical
with the standard Higgs decay $H_{SM} \rightarrow f \bar{f}$ contributions.
Weak MSSM contributions give sizeable contributions to the partial decay
width of the neutral light and heavy Higgs boson in the intermediate
pseudoscalar Higgs mass range $M_A$.
Vertex corrections with virtual gluino exchange are discussed separatly
for the decay channel $h^0 (H^0, A^0) \rightarrow f \bar{f}$.
Finally the branching ratios $h^0 \rightarrow b \bar{b}, c \bar{c},
\tau^+ \tau^-$ are presented.
\smallskip
\subsubsection{QED and QCD corrections}
\vspace*{0.1cm} \hspace*{0.5cm}
QED/QCD corrections to the partial decay width
$H \rightarrow f \bar{f}$ appear through vertex diagrams with
virtual photons/gluons and final state photon/gluon radiation \cite{braaten,
bardin,Drees2}:
\begin{equation}
 \Delta\Gamma_{QED} =\Delta\Gamma_{QED}^V +\Delta\Gamma_{QED}^B
                    = \Gamma_1 \cdot \delta_{QED} \ ,
\end{equation}
where $\delta_{QED}$ for the scalar Higgs bosons reads:
\begin{eqnarray}
\delta_{QED}^S & = & \frac{\alpha}{\pi} Q_f^2 \, \Delta_S \nonumber \\
\Delta_S & = &
 \frac{A (\tilde{\beta})}{\tilde{\beta}}
 + \frac{3 + 34 \tilde{\beta}^2 - 13 \tilde{\beta}^4}{16 \tilde{\beta}^3}
 \log \frac
 {1 + \tilde{\beta}}{1 - \tilde{\beta}} + \frac{3 ( -1 + 7 \tilde{\beta}^2 ) }
 { 8 \tilde{\beta}^2} \ .
\label{glqedk}
\end{eqnarray}
The pseudoscalar Higgs decay width $A \rightarrow f \bar{f}$ receives
a QED correction $\delta_{QED}$ different from Eq. (\ref{glqedk}),
\cite{Drees2} :
\begin{eqnarray}
\delta_{QED}^P & = & \frac{\alpha}{\pi} Q_f^2 \, \Delta_P \nonumber \\
\Delta_P & = &
\frac{A (\tilde{\beta})}{\tilde{\beta}}
 + \frac{19 + 2 \tilde{\beta}^2 + 3 \tilde{\beta}^4}{16 \tilde{\beta}}
 \log \frac
 {1 + \tilde{\beta}}{1 - \tilde{\beta}} + \frac{3 ( 7 - \tilde{\beta}^2 ) }
 { 8 } \ ,
\label{glqeda}
\end{eqnarray}
where
\begin{eqnarray}
 A (\tilde{\beta}) & = & (1+ \tilde{\beta}^2) \cdot \nonumber \\
  &  & \cdot \left[ 4 \, Li_2 (\frac{1-\tilde{\beta}}{1+\tilde{\beta}})
 + 2\, Li_2 (-\frac{1-\tilde{\beta}}{1+\tilde{\beta}})-3\log\frac{2}{1+
 \tilde{\beta}}
 \log\frac{1+\tilde{\beta}}{1-\tilde{\beta}}-2\log\tilde{\beta}\log
 \frac{1+\tilde{\beta}}
 {1 - \tilde{\beta}} \right] \nonumber \\
 & & - 3 \tilde{\beta} \log \frac{4}{1 - \tilde{\beta}^2} - 4 \tilde{\beta}
 \log \tilde{\beta} \ .
\end{eqnarray}
\smallskip \par
In the region $m_f^2 \ll M_H^2$ the QED correction for both scalar and
pseudoscalar Higgs particles, sufficiently accurate, is given by:
\begin{equation}
\delta_{QED} \simeq \frac{\alpha}{\pi} Q_f^2 \ [-3\, \log(\frac
{M_H}{m_f} ) + \frac{9}{4} ] \ .
\label{glappqed}
\end{equation}
\smallskip \par
The renormalization group improved QCD corrected decay width reads
\cite{braaten}:
\begin{equation}
\Gamma_{1, QCD} = \Gamma_{1} \cdot \frac{m_q^2 (M_H^2)}{m_{q,0}^2} \,
\left[ 1 + \frac{\alpha_s (M_H^2)}{\pi} C_F \, \left( \Delta_{S,P} +
3 \log \frac{M_H}{m_{q,0}} \right)
\, \right] \ ,
\end{equation}
where $m_q (M_H^2)$ is the effective quark mass from the
renormalization group equation \cite{jones}
\begin{eqnarray}
 m_q (q^2) & = & m_{q,0} \left( \frac{\beta_0 \alpha_s (q^2)}{2 \pi}
 \right)^{-\gamma_0 /2\beta_0 }
 \left[  1 + \frac{\beta_1 \gamma_0 - \beta_0 \gamma_1}{\beta_0^2}
 \frac{\alpha_s (q^2)}{8 \pi} \right. \nonumber \\ & & \left.       +
 \left( \frac{ (\beta_1 \gamma_0 - \beta_0 \gamma_1)^2}{2 \beta^4_0}
 + \frac{ \gamma_0 ( \beta_2 \beta_0 - \beta_1^2)}{\beta^3_0} +
 \frac{\gamma_1 \beta_1}{\beta^2_0} - \frac{\gamma_2}{\beta_0} \right)
 \, \left( \frac{\alpha_s (q^2)}{8 \pi} \right)^2 + ... \right] \ ,
 \nonumber \\
\label{renmass}
\end{eqnarray}
and $m_q (m_{q,0}^2) = m_{q,0}$ is the on-shell mass. The coefficients
in Eq. (\ref{renmass}) are:
\begin{eqnarray}
\beta_0 & = & \frac{33 - 2 N_f}{3} \nonumber \\
\beta_1 & = & 102 - \frac{38}{3} N_f \nonumber \\
\beta_2 & = & \frac{2857}{2} - \frac{5033}{18} N_f + \frac{325}{54} N_f^2
\nonumber \\[0.3cm]
\gamma_0 & = & - 8 \nonumber \\[0.1cm]
\gamma_1 & = & -\frac{404}{3} + \frac{40}{9} N_f \nonumber \\
\gamma_2 & = & \frac{2}{3} \left[ \frac{140}{27} N_f^2 +
\left( 160 \, \zeta (3) + \frac{2216}{9} \right) N_f - 3747 \right] \ ,
\end{eqnarray}
where $\zeta (3) \approx 1.2020596...$ and  $N_f = 6$ for $q^2 > 4 m_t^2$.
The strong coupling constant $\alpha_s$ with three-loop contributions
is given by:
\begin{eqnarray}
 \alpha_s (q^2) & = & \frac{4 \pi}{\beta_0 L_q} \left[
 1 - \frac{\beta_1}{\beta_0^2} \frac{\log L_q}{L_q} +
 \frac{\beta_1^2}{\beta_0^4} \frac{\log^2 L_q}{L_q^2} -
 \frac{\beta_1^2}{\beta_0^4} \frac{\log L_q}{L_q^2} +
 \frac{\beta_2 \beta_0 - \beta_1^2}{\beta_0^4} \frac{1}{L_q^2} \right] \ ,
\end{eqnarray}
where  $ L_q = \log (q^2/\Lambda_{QCD,N_f}^2) $ and $\Lambda_{QCD}^{N_f=5}
 = 150$ MeV. $\Lambda_{QCD}^{N_f=6}$ follows from the condition:
\begin{equation}
 \alpha_s^{(5)} (m_t) = \alpha_s^{(6)} (m_t) \ .
\end{equation}
The QED corrections are small for $b$-quarks ($-\delta_{QED} < 0.4 \%$ for
$M_H$ up to $1$ TeV) and somewhat bigger for $c$-quarks ($ < 1.8 \%$) and
$\tau$-leptons ($ < 4 \%$ ).
QED contributions to
$top$ quark decays are $\leq 0.4 \%$ for $M_H > 500$ GeV.
Near the top production threshold,
the Coulomb singularity appears and
non-perturbative effects have to be taken into account
(we do not consider these subtleties here).
The fermionic decay width for a scalar Higgs vanishes for $M_H \rightarrow
2 m_t$
\cite{Drees2}, whereas for the pseudoscalar Higgs a finite
contribution remains.
\smallskip \par
The QCD corrections to the hadronic decay width in terms of the
on-shell masses are large: in the $b \bar{b} \ (c \bar{c})$ channel,
$\delta_{QCD} = -39 \%$ for a light Higgs $<2 M_W$, increasing up
to $-60 \%$ $(-75\%)$ for $ 1$ TeV Higgs boson. The dominant part can
be absorbed in the running quark masses. For $M_H$ sufficiently above the
$t \bar{t}$ threshold, the QCD corrections to the $ H \rightarrow t \bar{t}$
width are typically $\delta_{QCD} \leq 15 \% $.
\smallskip
\subsubsection{Weak MSSM corrections}
\vspace*{0.1cm} \hspace*{0.5cm}
The partial decay width $h^0 \rightarrow b \bar{b}$ is the dominant decay
channel for a light scalar MSSM Higgs and a standard Higgs with a mass below
$M_{H} < 140$ GeV. A precise prediction of the $H \rightarrow b \bar{b}$
decay width and the fermionic $b \bar{b}$, $c \bar{c}$, $\tau^+ \tau^-$
branching ratios requires the inclusion of radiative corrections at
the one-loop level. One-loop contributions to the
$b \bar{b}$ decay channel are discussed in detail for the light (heavy) scalar
and pseudoscalar Higgs decay width $h^0 (H^0, A^0) \rightarrow b \bar{b}$.
The $c \bar{c}$, $\tau^+ \tau^-$ decay
channels are presented within the context of the Higgs branching ratios in
the following subsection.
\par \smallskip
The one-loop contributions from the residual MSSM particles are contained
in the electroweak decay width $\Gamma_1$, Eq. (\ref{gamma1}).
Figs. 2 a,b show the one-loop decay width $\Gamma_1$ for the light
Higgs decay channel $h^0 \rightarrow b \bar{b}$ as a function of the
light Higgs mass $M_{h^0}$. In Fig. 2 a, $\tan \beta = 2, 30$ while
$\tan \beta = 0.5, 8$ values are presented in Fig. 2 b.
Soft breaking parameters are $m_{sf} = 700$ GeV, $\mu = 100$ GeV,
$M = 550$ GeV and no mixing of left-right sfermion
states is assumed. The sfermion mass matrix is given in Eq. (\ref{sqmatrix}).
Appendix B contains the chargino and neutralino mass matrix.
The $h^0 \rightarrow b \bar{b}$
decay width dependence on the top-quark mass $m_t$ is shown for
$m_t = 160$ GeV (dotted line), $m_t = 175$ GeV (solid line) and
$m_t = 190$ GeV (dashed line), as favoured by the CDF data for the top-quark
mass \cite{top}.
In Figs. 2 a,b the upper limit of the light Higgs mass $M_{h^0}$ increases
$\sim  m_t^4$, as discussed in Eq. (\ref{glmapp2}). For a light Higgs mass
$M_{h^0} < 80 $ GeV and $\tan \beta \ge 5$ the partial decay width
$h^0 \rightarrow b \bar{b}$ is almost insensitive on $m_t$.
$\Gamma_1 (h^0 \rightarrow b \bar{b})$ increases with $ \sim m_t^4$ for a
light Higgs mass $M_{h^0} > 80$ GeV. This $m_t^4$ dependence of the
partial decay width is a universal contribution of the external Higgs
two-point functions, described by the one-loop mixing angle
$\sin^2 \alpha_{eff}$, in Eq. (\ref{glalpheffx}).
Figs. 2 c,d show the partial decay width as a function of the pseudoscalar
Higgs mass $M_A$. The decay width $\Gamma_1 (h^0 \rightarrow b \bar{b})$
reaches a maximum near $90$ GeV $< M_A <$ $ 110$ GeV and decreases for
$M_A > 110$ GeV. Figs. 2 a-d also show the standard Higgs partial decay
width $H_{SM} \rightarrow b \bar{b}$ with one-loop weak corrections
\cite{bardin,dabh}. The mass $M_{H_{SM}}$ is chosen to be
equal the light MSSM Higgs mass $M_{h^0}$. In Figs. 2 c,d the solid
(dashed) standard Higgs decay width corresponds to the MSSM $\tan \beta$
values $30$ ($2$) in Fig. 2 c and $8$ ($0.5$) in Fig. 2 d.
As a result, $\Gamma_1 (h^0 \rightarrow b \bar{b})$ in the MSSM is enhanced
for all $\tan \beta$ values compared to the standard decay width.
For large pseudoscalar masses $M_A \rightarrow \infty$, however, the MSSM
decay width approaches the standard model result closely. This
behaviour is discussed in terms of Eq. (\ref{glsina1}), where
$\sin^2 \alpha \rightarrow \cos^2 \beta$ in the limit $M_A \rightarrow \infty$.
In Fig. 2 the pseudoscalar Higgs mass range is chosen up to $300$ GeV.
The gap between
the MSSM and the standard decay width is sizeable for $M_A = 300$ GeV.
Larger pseudoscalar masses tend to approach the standard model decay
width as can be seen in Figs. 2 c,d.
Even in the limit of a large pseudoscalar mass $M_A$, the genuine vertex
corrections to the MSSM (SM) Higgs decay width are different, due to the
presence of virtual supersymmetric particles in Eq. (\ref{gamma1}).
\par \smallskip
The sfermion mass dependence on $\Gamma_1 (h^0 \rightarrow b \bar{b})$ is
shown in Figs 3 a,b as a function of the light Higgs mass $M_{h^0}$
and for $\tan \beta$ values $\tan \beta = 2, 30$ (Fig. 3 a),
$\tan \beta = 0.5, 8$ (Fig. 3 b). No mixing of left-right sfermion states
is assumed. In Figs. 3 a,b the sfermion soft breaking parameters are
$m_{sf} = 1$ TeV (solid line),$m_{sf} = 500$ GeV (dotted),
$m_{sf} = 300$ GeV (short dashed) and $m_{sf} = 200$ GeV (long dashed).
The upper limit of the light Higgs mass and the partial decay width
$\Gamma_1 (h^0 \rightarrow b \bar{b})$ increases
$\sim \log (\frac{ m_{\tilde{t}_L}  m_{\tilde{t}_R}}{m_t^2})$
, Eq. (\ref{glmapp2}). Mixing
effects from the left and right sfermions states are shown in Fig. 3 c for
$\tan \beta = 2, 30$. The off-diagonal mixing parameter $A_t'$ in Eq.
(\ref{glaprime}) is $A_t' = 0$, $100$ GeV, $200$ GeV, $300$ GeV, $400$ GeV.
These mixings increase the light Higgs $M_{h^0}$ and the partial
decay width $\Gamma_1 (h^0 \rightarrow b \bar{b})$ simultaneously.
\par \smallskip
Effects from gaugino soft breaking parameters $M$, $\mu$ are displayed
in Fig. 3 d. The dependence of the partial decay width
$\Gamma_1 (h^0 \rightarrow b \bar{b})$ on $M$ is shown for $M = 100, 200$
GeV to be very small.
$\mu$ enters the sfermion mass matrix, Eq. (\ref{sqmatrix}),
in the off-diagonal
entries and in the Higgs-sfermion couplings. In Fig. 3 d no left-right
mixing is present by fine-tuning the $A$ parameter. Large $\tan \beta$
values ($\tan \beta \ge 30$), however, show sizeable effects for the
partial decay width, since the parameter $A$ increases with $\tan \beta$,
$\mu$ and contributes to the $H$-$\tilde{f}_L$-$\tilde{f}_R$ couplings.
For lower $\tan \beta$ values the partial decay width is almost insensitive
on $\mu$.
\par \smallskip
Figs. 4 a,b show the one-loop partial decay width $\Gamma_1$ of the
heavy Higgs boson $H^0 \rightarrow b \bar{b}$ as a function of the
heavy Higgs mass $M_{H^0}$ for values $\tan \beta = 0.5, 2, 8, 30$.
The top quark mass dependence of the decay width $\Gamma_1 (H \rightarrow
b \bar{b})$ is presented in Fig. 4 a for $m_t = 160$ GeV (dotted line),
$175$ GeV (solid) and $190$ GeV (dashed), while Figs. 4 b shows the
decay width for several sfermion soft breaking parameters
$m_{sf} = 1$ TeV (solid line), $500$ GeV (dotted), $300$ GeV (short dashed)
and $200$ GeV (long dashed). The soft breaking parameters are described
in the figure caption. For a heavy scalar Higgs mass $M_{H^0} > 180$ GeV
and $\tan \beta > 2$ the decay width $\Gamma_1$ is almost insensitive on
$m_t$. $\Gamma_1 (H^0 \rightarrow b \bar{b})$ increases for large
$\tan \beta$ values.
\newpage
The pseudoscalar one-loop decay width $\Gamma_1 (A^0 \rightarrow b \bar{b})$
is shown in Fig. 4 c for fixed soft breaking parameters and  $\tan \beta$
values. The partial decay width $\Gamma_1 (A^0 \rightarrow f \bar{f})$ for
down(up) type fermions increases (decreases) with $\sim \tan^2 \beta \ (\cot^2
\beta)$. Vertex corrections $\delta \Gamma_b = 2 \Re e \Delta T_{A^0}$ in Eq.
(\ref{gamma1}) are shown in Fig. 4 d as a
function of the pseudoscalar mass $M_A$. The top quark mass dependence
of the vertex corrections is presented for $m_t = 160$, $175$, $190$ GeV
(dotted, solid, dashed line). The one-loop contributions $\delta \Gamma_b$
are large $\approx 15 \%$ for $\tan \beta = 0.5$. For $\tan \beta \ge 2$,
$\delta \Gamma_b \approx 3 - 8 \% $ and decreases with larger pseudoscalar
masses $M_A$.
\smallskip
\subsubsection{Vertex corrections of virtual gluinos}
\vspace*{0.1cm} \hspace*{0.5cm}
The supersymmetric partners of the $SU(3)$ gluons, the gluinos
$\tilde{g}_a$, appear as virtual states in the $H \rightarrow q \bar{q}$
vertex corrections (together with squarks) with the strong coupling
constant $\alpha_s$.
They contribute a shift $\delta \Gamma_{Gl}$ in the decay width for
$h^0, H^0, A^0 \rightarrow q \bar{q}$.
\begin{equation}
 \Gamma_{1,\, Gl} (H \rightarrow q \bar{q}) =
\Gamma_1 (H \rightarrow q \bar{q}) \ ( 1 + \delta \Gamma_{Gl} ) \ ,
\end{equation}
where
\begin{equation}
 \delta \Gamma_{Gl} = 2 \Re e \ ( \Delta T^{H}_{Gl} + \Sigma_{S, Gl}^f
 (m_q^2) - 2 m_q^2 ( \Sigma_{S, Gl}^{f'} (m_q^2) + \Sigma_{V, Gl}^{f'}
 (m_q^2) ) \ ) \ ,
\label{vgl}
\end{equation}
is the one-loop gluino contribution to the decay width
$H \rightarrow q \bar{q}$.
The vertex correction $\Delta T^{H}_{Gl}$ and self energies
$\Sigma_{S, Gl}^f$, $\Sigma_{V, Gl}^f$ are given in
Eq. (\ref{glcorr}, \ref{siggl}).
\par \smallskip
Mixing effects from virtual squarks in
the vertex corrections $\Delta T^{H}_{Gl}$ and quarks self energies
$\Sigma_{Gl}^{f}$ are described by a $2 \times 2$ squark mass matrix:
\begin{equation}
 \cal M_{\rm \tilde{q}}^{\rm 2} \rm = \left( \begin{array}{ll}
 M_{\tilde{Q}}^2 + m_q^2 + M_Z^2 (I_3 - Q_q s_W^2) \cos 2 \beta &
 m_q (A_q + \mu \{ \cot \beta , \tan \beta \} ) \\
 m_q (A_q + \mu \{ \cot \beta , \tan \beta \} ) &
 M_{\{\tilde{U},\tilde{D}\}}^2 + m_q^2 + M_Z^2 Q_q s_W^2 \cos 2 \beta
 \end{array} \right)
\label{sqmatrix}
\end{equation}
with SUSY soft breaking parameters $M_{\tilde{Q}}$,
$M_{\tilde{U}, \tilde{D}}$, $A_q$, and $\mu$. The notation in the
off-diagonal entries in Eq. (\ref{sqmatrix}):
\begin{equation}
A_q' = A_q + \mu \{ \cot \beta , \tan \beta \}
\label{glaprime}
\end{equation}
will be used.
In the following discussion the soft breaking parameters are taken to
be equal $m_{sf} = M_{\tilde{Q}} = M_{\tilde{U}, \tilde{D}}$.
Up and down type squarks in (\ref{sqmatrix}) are distinguished by
setting f=u,d and the $\{u,d\}$ entries in the parenthesis.
The parameter $\mu$ in the off-diagonal matrix elements in (\ref{sqmatrix})
is also present in the gaugino sector. The sfermion masses, obtained from
diagonalizing (\ref{sqmatrix}) are:
\begin{equation}
 m_{\tilde{q}_i}^2 = \frac{1}{2} (\rm Tr \cal M_{\rm \tilde{q}}^{\rm 2} \rm
 \pm \sqrt{ ( Tr \cal M_{\rm \tilde{q}}^{\rm 2} \rm ) ^2 - 4
 Det \cal M_{\rm \tilde{q}}^{\rm 2} \rm } \ ) \ , \ i=1,2 \ ,
\end{equation}
where the corresponding rotation matrices
\begin{equation}
 U (\theta_{\tilde{q}}) = \left( \begin{array}{rl}
  \cos \theta_{\tilde{q}} & \sin \theta_{\tilde{q}} \\
  - \sin \theta_{\tilde{q}} & \cos \theta_{\tilde{q}} \\
\end{array} \right) \ ,
\end{equation}
are described by the sfermion mixing angle  $\theta_{\tilde{q}}$:
\begin{equation}
 \tan 2 \theta_{\tilde{q}} = \frac{2 m_q ( A_q + \mu \{ \cot \beta,
 \tan \beta \} ) }{ M_{\tilde{q}}^2 - M_{\{\tilde{U},\tilde{D}\}}^2
 + M_Z^2 (I_3 - 2 Q_q s_W^2) \cos 2 \beta } \ .
\end{equation}
\par \smallskip
The one-loop vertex correction for the scalar neutral Higgs decay
$h^0 (H^0) \rightarrow q \bar{q} $ is given by:
\begin{equation}
 \Delta T^{h^0 (H^0)}_{Gl} = - \frac{ \alpha_s }{ 3 \pi} \ \sum_{i,j = 1}^{2}
 \ \frac{T_{i,j}^{h^0 (H^0)}}{T_{h^0(H^0)ff}} \ [ \ 2 m_q \ \delta_{i j}
 \ C_1^+ -  m_{\tilde{gl}} \ \Delta_{i j} \
 C_0 \ ] (M_{h^0(H^0)}^2, m_{\tilde{q}_i}, m_{\tilde{q}_j}, m_{
 \tilde{gl}} ) \ ,
\label{glcorr}
\end{equation}
where $T_{h^0(H^0) ff}$ in Eq. (\ref{glcorr}) are the tree-level couplings,
$\delta_{i j}$ is the unit matrix and
$$
 \Delta_{i j} =
 \left( \begin{array}{lr} \sin 2 \theta_{\tilde{q}} &
 \cos 2 \theta_{\tilde{q}} \\ \cos 2 \theta_{\tilde{q}} & -
 \sin 2 \theta_{\tilde{q}} \end{array} \right) \ .
$$
The pseudoscalar Higgs decay $ A^0 \rightarrow q \bar{q}$ yields
the vertex correction:
\begin{equation}
 \Delta T^{A^0} = - \frac{ \alpha_s }{ 3 \pi} \ \sum_{i,j = 1}^{2} \
 \frac{T_{i,j}^{A^0}}{T_{A^0 ff}}  \ \epsilon_{i j} \
 m_{\tilde{gl}} \ C_0 \ (M_{A^0}^2, m_{\tilde{q}_i}, m_{\tilde{q}_j}, m_{
 \tilde{gl}} ) \ \gamma_5 \ ,
\label{gpcorr}
\end{equation}
with $T_{A^0 ff}$ in Eq. (\ref{gpcorr}) from Tab. 2 and $ \epsilon_{12} =
- \epsilon_{21} = 1$.
$T_{i,j}^H$ are the $H$-$\tilde{q}_i$-$\tilde{q}_j$ couplings in the
squark mass eigenstate fields, obtained by the transformation:
\begin{equation}
 T_{i,j}^H = U (\theta_{\tilde{q}})_{i,a} \ T_{a,b}^H \
 U^\dagger (\theta_{\tilde{q}})_{b,j} \ ,
\end{equation}
where
\footnotesize
\begin{eqnarray}
 T_{a,b}^{h^0} & = & ig \left( \begin{array}{ll}
 \frac{M_Z}{c_W} (I_3 - Q_q s_W^2) \sin (\alpha + \beta) -
 \frac{m_q^2 \{ \cos \alpha,-\sin \alpha \}}{M_W \{ \sin \beta, \cos \beta
 \} } & \frac{m_q ( \mu \{ \sin
 \alpha,-\cos \alpha ) - A_q \{ \cos \alpha ,-\sin \alpha \} ) }{2 M_W \{
 \sin \beta , \cos \beta \} } \\
 \frac{m_q ( \mu \{ \sin
 \alpha,-\cos \alpha ) - A_q \{ \cos \alpha ,-\sin \alpha \} ) }{2 M_W \{
 \sin \beta , \cos \beta \} } &
 \frac{M_Z}{c_W} Q_q s_W^2 \sin (\alpha + \beta) -
 \frac{m_q^2 \{ \cos \alpha,-\sin \alpha \}}{M_W \{ \sin \beta, \cos \beta
 \} } \end{array} \right)
\nonumber \\
T_{a,b}^{H^0} & = & -ig \left( \begin{array}{ll}
 \frac{M_Z}{c_W} (I_3 - Q_q s_W^2) \cos (\alpha + \beta) +
 \frac{m_q^2 \{ \sin \alpha, \cos \alpha \}}{M_W \{ \sin \beta, \cos \beta
 \} } & \frac{m_q ( \mu \{ \cos
 \alpha, \sin \alpha ) + A_q \{ \sin \alpha , \cos \alpha \} ) }{2 M_W \{
 \sin \beta , \cos \beta \} } \\
 \frac{m_q ( \mu \{ \cos
 \alpha, \sin \alpha ) + A_q \{ \sin \alpha , \cos \alpha \} ) }{2 M_W \{
 \sin \beta , \cos \beta \} } &
 \frac{M_Z}{c_W} Q_q s_W^2 \cos (\alpha + \beta) +
 \frac{m_q^2 \{ \sin \alpha, \cos \alpha \}}{M_W \{ \sin \beta, \cos \beta
 \} } \end{array} \right) \nonumber \\
 T_{a,b}^{A^0} & = & g \left( \begin{array}{ll}
 0 & -\frac{m_q}{2 M_W} ( \mu - A_q \{ \cot \beta, \tan \beta \} ) \\
 \frac{m_q}{2 M_W} ( \mu - A_q \{ \cot \beta, \tan \beta \} ) & 0
 \end{array} \right) \ . \nonumber \\
\nonumber \\
\end{eqnarray}
\normalsize
The first (second) column in the parentheses belongs to the up (down)
squark coupling.
In Eq. (\ref{vgl}) the fermion self energy with virtual gluinos and squarks
for the scalar (vector) components reads:
\begin{eqnarray}
 \Sigma_S^f (m_q^2) & = & \frac{- \alpha_s}{3 \pi} \ \frac{m_{\tilde{gl}}}
 {m_q} \ \sin 2 \theta_{\tilde{q}} \ ( \ B_0
 (m_q^2, m_{\tilde{gl}}, m_{\tilde{q}_1} )
 - B_0 (m_q^2, m_{\tilde{gl}}, m_{\tilde{q}_2}) ) \nonumber \\
 \Sigma_{V, i}^f (m_q^2) & = & \frac{- \alpha_s}{3 \pi} \ \sum_{i=1}^2  \
 B_1 (m_q^2, m_{\tilde{gl}},
 m_{\tilde{q}_i} ) \ .
\label{siggl}
\end{eqnarray}
The integrals $ B_0$, $B_1, C_0, C_1$ are defined in appendix A.
\par \smallskip
The numerical analysis of the gluino contributions $\delta \Gamma_{Gl}$
is shown in Fig. 5. Light (heavy) scalar and pseudoscalar Higgs boson
decays $H \rightarrow b \bar{b}$ are presented in Figs. 5 (a,b), (c,d), (e,f).
In Figs. 5 a,c,e the one-loop corrections are shown as a function of the
Higgs mass and for two fixed gluino mass parameters $m_{\tilde{gl}} =
500$ GeV (solid line), $200$ GeV (dotted line). The sfermion mass is
$m_{sf} = 700$ GeV (solid line),
$m_{sf} = 500$ GeV (dotted line) and no left-right mixing is present.
The contribution $\delta \Gamma_{Gl}$ decreases for larger gluino masses
$m_{\tilde{gl}}$ and larger Higgs masses $M_H$. The corrections are
sizeable $\simeq 30 \%$ for large $\tan \beta = 30$ values and a light
mass $M_{h^0}$ below the upper mass limit.
Mixing effects from left-right sfermion states are shown in Fig. 5 b,d,f
as a function of the $\mu$ parameter and all other parameters fixed.
The corrections $\delta \Gamma_{Gl}$ are $\sim \mu$ and become large
($20 \%$) for lighter Higgs masses and $| \mu | \ge 250$ GeV.
\par \smallskip
\subsubsection{Fermionic branching ratios $H \rightarrow f \bar{f}$}
\vspace*{0.1cm} \hspace*{0.5cm}
Branching ratios of the fermionic Higgs decay channels are experimentally
measurable, even if the partial decay width $\Gamma (H\rightarrow f \bar{f})$
can not be measured directly.
In the following the branching ratios of the light neutral scalar MSSM
Higgs $h^0$ and the standard Higgs $H_{SM}$ in $b \bar{b}$, $c \bar{c}$,
$\tau^+ \tau^-$ are presented.
Here we restrict the discussion to the fermionic branching ratio
$R_f$, given by:
\begin{equation}
R_f = \frac{ \Gamma_1 ( H \rightarrow f \bar{f}) }{ \sum_{f = \tau, c, b, t}
\Gamma_1 ( H \rightarrow f \bar{f}) } \ ,
\label{glratio}
\end{equation}
where the light fermion contributions are negligible.
\smallskip
Figs. 6 a,b  show the light neutral MSSM branching ratios $R_f$ for
$b \bar{b}$ (Fig. 6 a) and the $c \bar{c}$, $\tau^+ \tau^-$ decay channels
(Fig. 6 b), where the full one-loop contributions from section 4.2 are
included. No QED/QCD and gluino contributions are included in the figure.
In Fig. 6 the branching ratio $R_f$ is a function of the light Higgs mass
$M_{h^0}$ and values $\tan \beta = 0.5, 2, 8, 30$ are shown for a top-quark
mass $m_t = 175$ GeV and soft breaking parameters $m_{sf} = 700$ GeV,
$\mu = 100$ GeV, $M = 550$ GeV.
The $h^0 \rightarrow b \bar{b}$ decay rate is $87 - 95 \%$ for
$0.5 \le \tan \beta \le 30$. $R_f$ decreases for the $b$ and $\tau$ decay
channels (increases for $c$) near the upper limit of the light
MSSM Higgs mass. In the limit $M_A \rightarrow \infty$ the branching ratio
$R_f$ reaches the standard model result closely, as shown by the dotted
lines in Figs. 6 a,b. Deviations from the standard model result are model
dependent supersymmetric vertex contributions.
$b$ and $\tau$ decay ratios $R_f$ are between $0 - 9 \%$ and $4 - 6 \%$ in
the range $0.5 \le \tan \beta \le 30$.
\par \smallskip
The branching ratio $R_f$ in Eq. (\ref{glratio}), where the approximation
formulae Eq. (\ref{glalpheffx}) for the partial decay width is used instead:
\begin{equation}
 \Gamma_1' (H \rightarrow f \bar{f} ) =  \frac{N_C \,G_F \,m_f^2 }
 { 4 \sqrt{2} \pi  } M_{H} \tilde{\beta}^n \ |\kappa_{H, eff}^f|^2 \ ,
\label{ratapp}
\end{equation}
with $\kappa_{h^0, eff}^d = - \frac{\sin \alpha_{eff}}{\cos \beta} \ , ...$
yields a qualitative good prediction within
$0.1 \%$ for $\tan \beta \ge 2$ and $0.6 \%$ for $\tan \beta = 0.5$ compared
to the complete result. The approximate result Eq. (\ref{ratapp}) is plotted
by the dashed line.
In the ratio $R_f$, Eq. (\ref{glratio}), the universal contributions
$\Delta r$ and the vertex correction part $\delta v / v$ from the one-loop
decay width Eq. (\ref{gamma1}) cancel. Therefore the complete result for
the branching ratio $R_f$ and the approximation formulae, Eq. (\ref{ratapp})
are in good agreement.
\subsection{Conclusions}
\vspace*{0.1cm} \hspace*{0.5cm}
The fermionic partial decay width $\Gamma_1 ( H \rightarrow f \bar{f} )$ for
the neutral MSSM Higgs bosons $h^0$, $H^0$, $A^0$ is calculated with full
one-loop MSSM contributions for the decay channels $b \bar{b}$, $c \bar{c}$,
$\tau^+ \tau^-$. In the calculation, the renormalization scheme for the
supersymmetric Higgs sector \cite{dabx} was used.
The tree-level decay width for down (up) type fermions is enhanced
(suppressed) compared to the standard model Higgs decay width. One-loop
corrections in the pseudoscalar Higgs mass range $80$ GeV $\le M_A \le 110$
GeV give large corrections $\sim m_t^4$ to the decay width
$\Gamma_1 ( h^0 (H^0) \rightarrow f \bar{f} )$. The diagonalization of the
neutral scalar Higgs mass matrix, described by the  mixing
angle $\alpha_{1-loop}$, receives the dominant contributions from $top$
and $stop$ loops $\cal{O} \rm (m_t^4)$.
The mixing angle $\sin^2 \alpha$ is calculated with full one-loop
contributions ($\sin^2 \alpha_{1-loop}$), in the effective potential
approximation ($\sin^2 \alpha_{eff}$) and as a flavour dependent effective
mixing angle ($\sin^2 \alpha_{f}$). The mixing angles $\sin^2 \alpha_{1-loop}$,
$\sin^2 \alpha_{eff}$ and $\sin^2 \alpha_{f}$ are in agreement within $8 \%$.
For large pseudoscalar Higgs masses $M_A \rightarrow
\infty$  the decay width  $\Gamma_1 ( h^0 \rightarrow f \bar{f} )$
approaches the standard model result. In this limit, non-universal model
dependent one-loop contributions to the decay width $\Gamma_1$ can
distinguish between a standard and MSSM Higgs boson and depend in detail
on the chosen parameters. Virtual gluino vertex corrections give sizeable
contributions to the $H \rightarrow b \bar{b}$ decay width (branching ratios).
The branching ratios for the light neutral scalar Higgs decay $h^0
\rightarrow b \bar{b}$, $c \bar{c}$, $\tau^+ \tau^-$ are presented. The full
calculation and the approximation formulae Eq. (\ref{ratapp}) are in
agreement within $0.2 - 0.6 \%$.
\par
\medskip
\bf Acknowledgements.\\ \rm
\smallskip
I am grateful to W. Hollik for suggesting this topic and reading the
manuscript and to S. Pokorski for various discussions.
%
%
%
\newpage
\appendix
\renewcommand{\thesubsection}{A}
\renewcommand{\theequation}{A.\arabic{equation}}
\subsection{Vertex corrections and self energies}
\setcounter{equation}{0}\setcounter{footnote}{0}
\vspace*{0.1cm} \hspace*{0.4cm}
The Feynman rules of the minimal supersymmetric standard model are given
in \cite{hunter}.
All analytical formulae are calculated in the 't Hooft-Feynman gauge.
The two- and three-point functions $B_0$, $B_1$, $B_{22}$, $C_0$, $C_1$
and $C_2$ are defined at the end of appendix A.
$f'$ denotes the isospin partner for the external fermion $f$ in the same
isodoublet. \par \smallskip
$\bullet \ $
Scalar MSSM $h^0 (H^0) \rightarrow f \bar{f}$ vertex
corrections\footnote{The upper(lower) line in the parentheses is
the $h^0$ $(H^0) \rightarrow f \bar{f}$ vertex correction.}:
\begin{eqnarray}
 \Delta T_{1} & = &  V_1^s \ (k^2,m_f,m_{f},m_{f},M_{Z},v_f,a_f)
 \nonumber \\[0.1cm]
  & \mbox{with} & v_f = \frac{I_3^f - 2 s_W^2 Q_f}{2 s_W c_W}
 \ ,\ a_f = \frac{I_3^f}{2 s_W c_W} \nonumber \\[0.2cm] \nonumber
 \Delta T_{2} & = & \frac{m_{f'}}
  { 8 m_f s_W^2 } \frac{\kappa_{H}^{f'}}{\kappa_{H}^{f}} \
  V_1^s \ (k^2,m_f,m_{f'},m_{f'},M_{W},1,-1)
 \nonumber \\[0.2cm]
 \Delta T_{3} & = &  - \frac{ m_{f'} }{ 8 M_W^2 m_f s_W^2 }
 \frac{\kappa_{H}^{f'}}{\kappa_{H}^{f}} \
 V_2^s \ (k^2,m_f,m_{f'},m_{f'},m_{H^+},\lambda_f,\mu_f) \nonumber \\
 &  \mbox{with} & \lambda_f  = \left\{ \begin{array}{ll}
     m_f \tan \beta + m_{f'} \cot \beta & , \ f = d\\
     m_{f'} \tan \beta + m_f \cot \beta & , \ f = u\\
                \end{array} \right. \nonumber \\
 & & \mu_f  = \left\{ \begin{array}{rl}
     - m_f \tan \beta + m_{f'} \cot \beta & , \ f = d\\
     + m_{f'} \tan \beta - m_f \cot \beta & , \ f = u\\
     \end{array} \right.
 \nonumber \\[0.2cm]  \nonumber
 \Delta T_{4} & = & - \frac{ m_{f'}}{ 8 M_W^2 m_f s_W^2 }
 \frac{\kappa_{H}^{f'}}{\kappa_{H}^{f}} \
 V_2^s \ (k^2,m_f,m_{f'},m_{f'},M_{W},\nu_f,\pi_f) \nonumber \\
 &  \mbox{with} & \nu_f = \left\{ \begin{array}{ll}
    m_{f'} - m_f & , \ f = d\\
    m_f - m_{f'} & , \ f = u\\
                \end{array} \right. \nonumber \\ \nonumber
 & & \pi_f = \left\{ \begin{array}{rl}
    m_{f'} + m_f  & , \ f = d\\
  - m_f - m_{f'} & , \ f = u\\
    \end{array} \right.
 \nonumber \\[0.2cm]  \nonumber
 \Delta T_{5} & = &  \frac{ m_{f}^2 (\kappa_{H^0}^{f})^2}
   { 4 M_W^2  s_W^2 } \ V_2^s \ (k^2,m_f,m_{f},m_{f},m_{H^0},1,0)
  \nonumber \\[0.3cm]
 \Delta T_{6} & = &  \frac{ m_{f}^2 (\kappa_{h^0}^f)^2 }
   { 4 M_W^2  s_W^2  } \ V_2^s \ (k^2,m_f,m_{f},m_{f},m_{h^0},1,0)
 \nonumber \\[0.2cm]
 \Delta T_{7} & = & - \frac{ m_{f}^2 |\kappa_{A^0}^f|^{2} }
   { 4 M_W^2  s_W^2  } \ V_2^s \ (k^2,m_f,m_{f},m_{f},M_{A^0},0,1)
 \nonumber \\[0.2cm]
 \Delta T_{8} & = & - \frac{ m_{f}^2 }
   { 4 M_W^2  s_W^2  } \ V_2^s \ (k^2,m_f,m_{f},m_{f},M_{Z},0,1)
 \nonumber \\[0.2cm]
 \Delta T_{9} & = & \frac{ M_W}
 {m_f  s_W^2 \kappa_H^f} \, \sum_{i,j,k = 1}^{2,2,2}  V_3^s \
 (k^2,m_f,m_{\tilde{\chi}_j^+},m_{\tilde{\chi}_i^+},m_{\tilde{f}'_k},v_{jk}^f,
 a_{jk}^f,a_{ik}^f,v_{ik}^f,O_{ij}^{H},O_{ji}^{H}) \nonumber \\
 & \mbox{with} &
 v_{ik}^{d,u} = \left\{ \begin{array}{lr}
  (V_{i1},U_{i1}) \cos \tilde{\theta}_{f'} - \frac{m_{f'}}{\sqrt{2} M_W}
  \frac{V_{i2}, U_{i2}}{\sin \beta,\cos \beta} \sin \tilde{\theta}_{f'}
 & ,\ k=1 \nonumber \\
  (V_{i1},U_{i1}) \sin \tilde{\theta}_{f'} + \frac{m_{f'}}{\sqrt{2} M_W}
  \frac{V_{i2}, U_{i2}}{\sin \beta,\cos \beta} \cos \tilde{\theta}_{f'}
 & ,\ k=2 \nonumber \\
   \end{array} \right.  \\ & &
 a_{ik}^{d,u} = \left\{ \begin{array}{lr}
  - \frac{m_f}{\sqrt{2} M_W}
  \frac{U_{i2}, V_{i2}}{\cos \beta,\sin \beta} \cos \tilde{\theta}_{f'}
 & ,\ k=1 \nonumber \\
  - \frac{m_f}{\sqrt{2} M_W}
  \frac{U_{i2}, V_{i2}}{\cos \beta,\sin \beta} \sin \tilde{\theta}_{f'}
 & ,\ k=2 \nonumber \\
   \end{array} \right. \\ & &
 O_{ij}^H = \frac{1}{\sqrt{2}} \left\{ \begin{array}{lr}
  V_{i1} U_{j2} \sin \alpha - V_{i2} U_{j1} \cos \alpha & ,\ H = h^0
   \nonumber \\
  V_{i1} U_{j2} \cos \alpha + V_{i2} U_{j1} \sin \alpha & ,\ H = H^0
  \nonumber \\ \end{array} \right.
 \nonumber \\[0.2cm]
 \Delta T_{10} & = & \frac{2 M_W}{m_f \kappa_H^f}\, \sum_{i,j,k = 1}^{4,4,2}
 V_3^s \, (k^2,m_f,m_{\tilde{\chi}_j^0},m_{\tilde{\chi}_i^0},m_{\tilde{f}_k},
 v'^f_{jk},a'^f_{jk},a'^f_{ik},v'^f_{ik},O_{ij}'^{H},O_{ji}'^{H})
 \nonumber
 \\[0.1cm]  & \mbox{with} &
 v'^{d,u}_{ik} = \left\{ \begin{array}{lr}
  (Q_f N'_{i1} \mp \frac{1/2 + Q_f s_W^2}{s_W c_W} N'_{i2})
  \cos \tilde{\theta}_{f} + \frac{m_f (N_{i3},N_{i4})}{ 2 M_W s_W
  (\cos \beta , \sin \beta ) }
  \sin \tilde{\theta}_{f} & ,\ k=1 \nonumber \\
  (Q_f N'_{i1} \mp \frac{1/2 + Q_f s_W^2}{s_W c_W} N'_{i2})
  \sin \tilde{\theta}_{f} - \frac{m_f (N_{i3},N_{i4})}{ 2 M_W s_W
  (\cos \beta \sin \beta ) }
  \cos \tilde{\theta}_{f} & ,\ k=1 \nonumber \\
  \end{array} \right. \nonumber \\ & &
 a'^{d,u}_{ik} = \left\{ \begin{array}{lr}
 \frac{m_f N_{i3} }{2 M_W s_W (\cos \beta , \sin \beta ) }
 \cos \tilde{\theta}_{f}
 - (Q_f N'_{i1} - \frac{  Q_f s_W}{c_W} N'_{i2}) \sin \tilde{\theta}_{f}
 & ,\ k=1 \nonumber \\
 \frac{m_f N_{i3} }{ 2 M_W s_W (\cos \beta , \sin \beta ) }
 \sin \tilde{\theta}_{f}
 + (Q_f N'_{i1} - \frac{  Q_f s_W}{c_W} N'_{i2} ) \cos \tilde{\theta}_{f}
 & ,\ k=2 \nonumber \\
 \end{array} \right. \nonumber \\ & &
 O'^H_{ij} = \frac{1}{2} \left\{ \begin{array}{lr}
 Q_{ij} \sin \alpha + S_{ij} \cos \alpha  & ,\ H=h^0 \nonumber \\
 Q_{ij} \cos \alpha - S_{ij} \sin \alpha  & ,\ H=H^0 \nonumber \\
 \end{array} \right. \nonumber \\ & &
 Q_{ij} = ( N_{i3}  (N_{j2} - N_{j1} \tan \theta_W ) + N_{j3} (N_{i2}
 - N_{i1} \tan \theta_W) )/2 \nonumber \\ & &
 S_{ij} = ( N_{i4}  (N_{j2} - N_{j1} \tan \theta_W ) + N_{j4} (N_{i2}
 - N_{i1} \tan \theta_W) )/2
\nonumber   \\[0.1cm]
 \Delta T_{11} & = & \frac{1}{4 m_f s_W^2 \kappa_H^f}
 \left\{ \begin{array}{r} - \cos (\beta - \alpha) \\
  \sin (\beta - \alpha) \end{array} \right\} \
 V_4^s \ (k^2,m_f,M_W,m_{H^+},m_{f'},1/2,-1/2,\lambda_f,\mu_f)
  \nonumber \\ \nonumber
\nonumber
\\[0.3cm] \nonumber
 \Delta T_{12} & = &  \frac{1}{2 s_W c_W \kappa_H^f}
 \left\{ \begin{array}{r} \cos (\beta - \alpha) \\
 - \sin (\beta - \alpha) \end{array} \right\} \
 V_4^s \ (k^2,m_f,M_Z,M_{A},m_{f},v_f,a_f,0,-1) \cdot
 \\ \nonumber
 & & \ \cdot \ \left\{ \begin{array}{ll}
  \tan \beta & ,\ f = d\\
  \cot \beta & ,\ f = u\\
 \end{array} \right. \\[0.3cm]  \nonumber
 \Delta T_{13} & = &  \frac{1}{4 m_f s_W^2 \kappa_H^f}
 \left\{ \begin{array}{r} \sin (\beta - \alpha) \\
  \cos (\beta - \alpha) \end{array} \right\} \
 V_4^s \ (k^2,m_f,M_W,M_{W},m_{f'},1/2,-1/2,\nu_f,\pi_f)
 \\ \nonumber
\\[0.3cm]  \nonumber
 \Delta T_{14} & = &  \frac{1}{2 s_W c_W \kappa_H^f}
 \left\{ \begin{array}{r} \sin (\beta - \alpha) \\
 \cos (\beta - \alpha) \end{array} \right\} \
 V_4^s \ (k^2,m_f,M_Z,M_Z,m_{f},v_f,a_f,0,-1) \cdot
 \\ \nonumber
 & & \ \cdot \  \left\{ \begin{array}{rl}
   1 & ,\ f = d\\
  -1 & ,\ f = u\\
 \end{array} \right. \\[0.3cm]  \nonumber
 \Delta T_{15} & = & - \frac{1}{ 4 m_f s_W^2 \kappa_H^f}
 \left\{ \begin{array}{r}
  \sin ( \beta - \alpha ) + \cos 2 \beta \sin ( \beta - \alpha ) /
 (2 c_W^2) \\
  \cos ( \beta - \alpha ) - \cos 2 \beta \cos ( \beta - \alpha ) /
 (2 c_W^2)  \end{array} \right\} \cdot \\ \nonumber
 & &  \cdot \ V_5^s
(k^2,m_f,m_{H^+},m_{H^+},m_{f'},\lambda_f,\mu_f,\lambda_f,-\mu_f)
 \\[0.2cm]  \nonumber
 \Delta T_{16} & = & \frac{m_f (\kappa_{h^0}^f)^2}{4 s_W^2 c_W^2 \kappa_H^f}
 \left\{ \begin{array}{r}
  3 \cos 2 \alpha \sin (\alpha + \beta ) \\
  2 \sin 2 \alpha \sin ( \beta + \alpha ) - \cos ( \beta + \alpha )
  \cos 2 \alpha  \end{array} \right\} \cdot \\ \nonumber & &
 \cdot \ V_5^s \ (k^2,m_f,m_{h^0},m_{h^0},m_{f},1,0,1,0)
 \\[0.3cm]  \nonumber
 \Delta T_{17} & = &  \frac{ m_f (\kappa_{H^0}^f )^2}{4 s_W^2 c_W^2
 \kappa_{H}^f} \left\{ \begin{array}{r}
 -2 \sin 2 \alpha \cos ( \beta + \alpha ) - \sin ( \beta + \alpha )
 \cos 2 \alpha ) \\ 3 \cos 2 \alpha \cos ( \beta + \alpha )
 \end{array} \right\} \cdot
 \\ \nonumber & & \cdot \ V_5^s \ (k^2,m_f,m_{H^0},m_{H^0},m_{f},1,0,1,0)
  \\[0.2cm]  \nonumber
 \Delta T_{18} & = &  \frac{m_f \sin \alpha \cos \alpha \, \kappa_{h^0}^f
 \kappa_{H^0}^f}{2 s_W^2 c_W^2 \kappa_H^f} \left\{ \begin{array}{r}
 (2 \sin 2 \alpha \sin ( \alpha + \beta ) - \cos ( \alpha + \beta )
 \cos 2 \alpha \ ) \\
 - ( 2 \sin 2 \alpha \cos ( \alpha + \beta ) + \sin ( \alpha + \beta )
 \cos 2 \alpha \ ) \end{array} \right\} \
 \cdot \\ \nonumber & &  \cdot \ V_5^s \,
 (k^2,m_f,m_{h^0},m_{H^0},m_{f},1,0,1,0)
 \\[0.2cm]  \nonumber
 \Delta T_{19} & = & - \frac{m_f \, \cos 2 \beta \, (\kappa_{A^0}^f)^2 }
 { 4 s_W^2 c_W^2 \kappa_H^f} \left\{ \begin{array}{r}
 \sin ( \beta + \alpha )  \\ - \cos ( \beta + \alpha )
 \end{array} \right\}
 \ V_5^s \ (k^2,m_f,M_{A},M_{A},m_{f},0,1,0,1)
 \\[0.2cm]  \nonumber
 \Delta T_{20} & = & \frac{ m_f \, \cos 2 \beta }
      { 4 s_W^2 c_W^2 \kappa_H^f}
 \left\{ \begin{array}{r}
 \sin ( \beta + \alpha )  \\ - \cos ( \beta + \alpha )
 \end{array} \right\} \
 V_5^s \ (k^2,m_f,M_{Z},M_{Z},m_{f},0,1,0,1)
 \\[0.4cm]  \nonumber
 \Delta T_{21} & = &  \frac{- \cos 2 \beta }
  { 8 m_f s_W^2 c_W^2 \kappa_H^f}
 \left\{ \begin{array}{r}
 - \sin ( \beta + \alpha ) \\
   \cos ( \beta + \alpha )
 \end{array} \right\} \,
 V_5^s \, (k^2,m_f,M_{W},M_{W},m_{f'},\nu_f,\pi_f,\nu_f,-\pi_f)
 \\[0.3cm]  \nonumber
 \Delta T_{22} & = & - \frac{ m_f \ \sin 2 \beta }
      { 2 s_W^2 c_W^2 \kappa_H^f}
 \left\{ \begin{array}{r}
 - \sin ( \beta + \alpha ) \\ \cos ( \beta + \alpha )
 \end{array} \right\} \ \cdot \\ \nonumber \\ & & \
 \cdot \ V_5(k^2,m_f,M_{Z},M_{A},m_{f},0,1,0,1) \cdot
 \left\{ \begin{array}{rl}
     \tan \beta & , \ f = d\\
    -\cot \beta & , \ f = u\\
 \end{array} \right. \nonumber \\[0.1cm]  \nonumber
 \\[0.1cm]
 \Delta T_{23} & = &  \frac{1}{4 M_W^2 m_f s_W^2 \kappa_H^f}
 \left\{ \begin{array}{r}
  - \cos ( \beta - \alpha ) \ ( m_{H^+}^2 - m_{h^0}^2 )  \\
    \sin ( \beta - \alpha ) \ ( m_{H^+}^2 - m_{H^0}^2 )
 \end{array} \right\} \cdot \nonumber \\[0.2cm] & & \cdot \
 V_5^s \ (k^2,m_f,M_{W},m_{H^+},m_{f'},\nu_f,\pi_f,\lambda_f,\mu_f)
 \nonumber \\[0.4cm]
 \Delta T_{24} & = &  \frac{M_W^2 }{4 m_f s_W^2 \kappa_H^f}
 \left\{ \begin{array}{r} \sin (\beta - \alpha) \\
 \cos (\beta - \alpha) \end{array} \right\} \
 V_6^s \ (k^2,m_f,M_W,M_W,m_{f'},1,-1,1,-1)
 \nonumber \\[0.2cm]  \nonumber
 \Delta T_{25} & = &  \frac{2 M_W^2 }{m_f c_W^2 \kappa_H^f}
 \left\{ \begin{array}{r} \sin (\beta - \alpha) \\
 \cos (\beta - \alpha)  \end{array} \right\} \
 V_6 \ (k^2,m_f,M_Z,M_Z,m_{f},v_f,a_f,v_f,a_f)
 \\[0.2cm]  \nonumber
 \Delta T_{26} & = & \frac{M_W}{2 s_W^2 m_f \kappa_H^f} \,
 \sum_{i=1}^{2} \ [ \,
 (\cos^2 \tilde{\theta}_{f'} \, u_{H,1} +
 \sin^2 \tilde{\theta}_{f'} \, u_{H,2} + \sin 2 \tilde{\theta}_{f'}
 \, u_{H,3} )  \nonumber \\ & & \cdot
 V_5^s \, (k^2,m_f,m_{\tilde{f'}_1},m_{\tilde{f'}_1},m_{\tilde{\chi}_i^+},
 v_{i1}^f + a_{i1}^f, v_{i1}^f-a_{i1}^f, v_{i1}^f-a_{i1}^f,
 v_{i1}^f+a_{i1}^f) \nonumber \\ & &
 + (\sin^2 \tilde{\theta}_{f'} \, u_{H,1} +
 \cos^2 \tilde{\theta}_{f'} \, u_{H,2} - \sin 2 \tilde{\theta}_{f'}
 \, u_{H,3} ) \nonumber \\ & & \cdot
 V_5^s \, (k^2,m_f,m_{\tilde{f'}_2},m_{\tilde{f'}_2},m_{\tilde{\chi}_i^+},
 v_{i2}^f+a_{i2}^f,v_{i2}^f-a_{i2}^f,v_{i2}^f-a_{i2}^f,
 v_{i2}^f+a_{i2}^f) \nonumber \\ & &
 + (\sin 2 \tilde{\theta}_{f'} \, (u_{H,2} - u_{H,1}) +
 2 \cos 2 \tilde{\theta}_{f'} \, u_{H,3} ) \nonumber \\ & & \cdot
 V_5^s \, (k^2,m_f,m_{\tilde{f'}_1},m_{\tilde{f'}_2},m_{\tilde{\chi}_i^+},
 v_{i1}^f+a_{i1}^f,v_{i1}^f-a_{i1}^f,v_{i2}^f-a_{i2}^f,v_{i2}^f+a_{i2}^f) \, ]
\nonumber \\[0.3cm]
 \Delta T_{27} & = & \frac{- M_W}{m_f \kappa_H^f} \, \sum_{i=1}^{4} \ [ \,
 (\cos^2 \tilde{\theta}_{f} \, u_{H,1} +
 \sin^2 \tilde{\theta}_{f} \, u_{H,2} + \sin 2 \tilde{\theta}_{f}
 \, u_{H,3} )  \nonumber \\ & & \cdot
 V_5^s \, (k^2,m_f,m_{\tilde{f}_1},m_{\tilde{f}_1},m_{\tilde{\chi}_i^0},
 v_{i1}'^f+a_{i1}'^f,v_{i1}'^f-a_{i1}'^f,v_{i1}'^f-a_{i1}'^f,
 v_{i1}'^f+a_{i1}'^f) \nonumber \\ & &
 + (\sin^2 \tilde{\theta}_{f} \, u_{H,1} +
 \cos^2 \tilde{\theta}_{f} \, u_{H,2} - \sin 2 \tilde{\theta}_{f}
 \, u_{H,3} ) \nonumber \\ & & \cdot
 V_5^s \, (k^2,m_f,m_{\tilde{f}_2},m_{\tilde{f}_2},m_{\tilde{\chi}_i^0},
 v_{i2}'^f+a_{i2}'^f,v_{i2}'^f-a_{i2}'^f,v_{i2}'^f-a_{i2}'^f,v_{i2}'^f
 +a_{i2}'^f) \nonumber \\ & &
 + (\sin 2 \tilde{\theta}_{f} \, (u_{H,2} - u_{H,1}) +
 2 \cos 2 \tilde{\theta}_{f} \, u_{H,3} ) \nonumber \\ & & \cdot
 V_5^s \, (k^2,m_f,m_{\tilde{f}_1},m_{\tilde{f}_2},m_{\tilde{\chi}_i^0},
 v_{i1}'^f+a_{i1}'^f,v_{i1}'^f-a_{i1}'^f,v_{i2}'^f-a_{i2}'^f,v_{i2}'^f
 +a_{i2}'^f) \, ] \nonumber \\
\end{eqnarray}
where
\begin{eqnarray}
 (u_{h^0 ,j}) & = & \left( \begin{array}{l}
 \frac{M_Z}{c_W} ( \pm \frac{1}{2} - Q_\pm s_W^2 )
 \sin ( \alpha + \beta ) - \frac{ m_\pm^2 \{ \cos \alpha , -
 \sin \alpha \} }{ M_W \{ \sin \beta , \cos \beta \} }  \\
 \frac{M_Z}{c_W} Q_\pm s_W^2
 \sin ( \alpha + \beta ) - \frac{ m_\pm^2 \{ \cos \alpha , -
 \sin \alpha \} }{ M_W \{ \sin \beta , \cos \beta \} }  \\
 \frac{- m_f}{2 M_W \{ \sin \beta, \cos \beta \} }
 \{ \mu \sin \alpha - A_u \cos \alpha, \
 \mu \cos \alpha - A_d \sin \alpha \}  \\
 \end{array} \right) \nonumber \\[0.3cm]
 (u_{H^0 ,j}) & = & \left( \begin{array}{l}
 - \frac{M_Z}{c_W} ( \pm \frac{1}{2} - Q_\pm s_W^2 )
  \cos( \alpha + \beta ) - \frac{ m_\pm^2 \{ \sin \alpha ,
 \cos \alpha \} }{ M_W \{ \sin \beta , \cos \beta \} } \\
 - \frac{M_Z}{c_W} Q_\pm s_W^2
 \cos( \alpha + \beta ) - \frac{ m_\pm^2 \{ \sin \alpha ,
 \cos \alpha \} }{ M_W \{ \sin \beta , \cos \beta \} }  \\
 \frac{- m_f}{2 M_W \{ \sin \beta, \cos \beta \} }
 \{ \mu \cos \alpha + A_u \sin \alpha , \
 \mu \sin \alpha + A_d \cos \alpha \}  \\
 \end{array} \right)
 \nonumber \\
\end{eqnarray}
\par \smallskip
$\bullet$
Pseudoscalar MSSM $A^0 \rightarrow f \bar{f}$ vertex correction:
\begin{eqnarray}
 \Delta T_{1} & = &  V_1^p (k^2,m_f,m_{f},m_{f},M_Z,v_f,a_f )
 \nonumber \\[0.2cm]
 \Delta T_{2} & = & \frac{m_{f'}}{8 m_f s_W^2}
 \ \frac{|\kappa_{A^0}^{f'}|}{|\kappa_{A^0}^f|}
 V_1^p (k^2,m_f,m_{f'},m_{f'},M_W,1,-1 )
 \nonumber \\[0.2cm]
 \Delta T_{3} & = & - \frac{m_{f'}}{8 M_W^2 m_f s_W^2} \
 \frac{|\kappa_{A^0}^{f'}|}{|\kappa_{A^0}^f|}
 V_2^p (k^2,m_f,m_{f'},m_{f'},m_{H^+},\lambda_f,\kappa_f )
 \nonumber \\[0.2cm]
 \Delta T_{4} & = & - \frac{m_{f'}}{8 M_W^2 m_f s_W^2} \
 \frac{|\kappa_{A^0}^{f'}|}{|\kappa_{A^0}^f|}
 V_2^p (k^2,m_f,m_{f'},m_{f'},M_W,\nu_f,\pi_f )
 \nonumber \\[0.2cm]
 \Delta T_{5} & = & \frac{ m_f^2 (\kappa_{H^0}^f )^2 }{4 M_W^2 s_W^2 } \
 V_2^p (k^2,m_f,m_f,m_f,m_{H^0},1,0)
 \nonumber \\[0.2cm]
 \Delta T_{6} & = & \frac{ m_f^2 (\kappa_{h^0}^f )^2 }{4 M_W^2 s_W^2 } \
 V_2^p (k^2,m_f,m_f,m_f,m_{h^0},1,0)
 \nonumber \\[0.2cm]
 \Delta T_{7} & = & - \frac{ m_f^2 |\kappa_{A^0}^f |^2 }{ 4 M_W^2 s_W^2 } \
 V_2^p (k^2,m_f,m_f,m_f,M_{A},0,1)
 \nonumber \\[0.2cm]
 \Delta T_{8} & = & - \frac{ m_f^2 }{ 4 M_W^2 s_W^2 } \
 V_2^p (k^2,m_f,m_f,m_f,M_Z,0,1)
 \nonumber \\[0.2cm]
 \Delta T_{9} & = & \frac{M_W}{m_f s_W^2 |\kappa_{A^0}^f|} \, \sum_{i,j,k = 1}
 ^{2,2,2} \ V_3^p \ (k^2,m_f,m_{\tilde{\chi}_j^+},m_{\tilde{\chi}_i^+},
 m_{\tilde{f}'_k},v_{jk}^f,a_{jk}^f,a_{ik}^f,v_{ik}^f,O_{ij}^A,Q_{ji}^A)
 \nonumber \\[0.1cm]
 & \mbox{with} &  O_{ij}^A = \frac{1}{\sqrt{2}} \, ( V_{i1} U_{j2} \sin \beta
  + V_{i2} U_{j1} \cos \beta \, )
 \nonumber \\[0.3cm]
 \Delta T_{10} & = &  \frac{2 M_W}{m_f |\kappa_{A^0}^f|}
 \ \sum_{i,j,j = 1}^{4,4,2} \ V_3^p \
 (k^2,m_f,m_{\tilde{\chi}_j^0},m_{\tilde{\chi}_i^0},m_{\tilde{f}_k},
 v_{jk}'^f,a_{jk}'^f,a_{ik}'^f,v_{ik}'^f,P^A_{ij},P^A_{ji}) \nonumber
 \\[0.1cm]
 & \mbox{with} & P^A_{ij} = \frac{1}{2} \, ( Q_{ij} \sin \beta - S_{ij}
 \cos \beta \, )
 \nonumber \\[0.3cm]
 \Delta T_{11} & = & \frac{1}{4 s_W^2 m_f |\kappa_{A^0}^f|} \
 V_4^p \ (k^2,m_f,M_W,m_{H^+},m_{f'},1/2,-1/2,\lambda_f,\mu_f )
 \nonumber \\[0.2cm]
 \Delta T_{12} & = & \frac{ \cos ( \beta - \alpha ) \kappa_{h^0}^f}
 {2 s_W c_W |\kappa_{A^0}^f| } \
 V_4^p \ (k^2,m_f,M_Z,m_{h^0},m_f,v_f,-a_f,1,0 )
 \nonumber \\[0.2cm]
 \Delta T_{13} & = & - \frac{ \sin ( \beta - \alpha ) \kappa_{H^0}^f}
 {2 s_W c_W |\kappa_{A^0}^f|} \
 V_4^p (k^2,m_f,M_Z,m_{H^0},m_f,v_f,-a_f,1,0 )
 \nonumber \\[0.2cm]
 \Delta T_{14} & = & \frac{m_f \cos 2 \beta \sin ( \beta + \alpha )
 \kappa_{h^0}^f}{2 s_W^2 c_W^2} \
 V_5^p \ (k^2,m_f,M_A,m_{h^0},m_f,0,1,1,0 )
 \nonumber \\[0.2cm]
 \Delta T_{15} & = & - \frac{m_f \cos 2 \beta \cos
 (\beta + \alpha ) \kappa_{H^0}^f }{2 s_W^2 c_W^2} \
 V_5^p \ (k^2,m_f,M_A,m_{H^0},m_f,0,1,1,0 )
 \nonumber \\[0.2cm]
 \Delta T_{16} & = & - \frac{m_f \sin 2 \beta \cos ( \beta + \alpha ) \,
 \kappa_{H^0}^f \, \kappa_{H^0}^{f'} }{2 s_W^2 c_W^2 \kappa_{h^0}^f } \
 V_5^p \ (k^2,m_f,M_Z,m_{H^0},m_f,0,1,1,0)
 \nonumber \\[0.2cm]
 \Delta T_{17} & = & \frac{m_f \sin 2 \beta \sin ( \beta + \alpha ) \,
 \kappa_{H^0}^{f'}}{2 s_W^2 c_W^2} \ V_5^p \ (k^2,m_f,M_Z,m_{h^0},m_f,0,1,1,0)
 \nonumber \\[0.2cm]
 \Delta T_{18} & = & - \frac{1}{2 s_W^2 m_f |\kappa_{A^0}^f| } \
 V_5^p \ (k^2,m_f,M_W,m_{H^+},m_{f'},\nu_f,\pi_f,\lambda_f,\mu_f )
 \nonumber \\[0.2cm]
 \Delta T_{19} & = & \frac{m_f'}{2 m_f s_W |\kappa_{A^0}^f| } \,
 \left\{ \begin{array}{rl}
 \mu - A_u \cot \beta  & , \ f = d \\
 \mu - A_d \tan \beta  & , \ f = u
 \end{array} \right\} \cdot  \nonumber \\ & &
 \cdot \, \sum_{i = 1}^2 \,
 V_{5}^p \, (k^2,m_f,m_{\tilde{f}'_1},m_{\tilde{f}'_2},m_{\tilde{\chi}_i^+},
 v_{i1}^f+a_{i1}^f,v_{i1}^f-a_{i1}^f,
 v_{i2}^f-a_{i2}^f,v_{i2}^f+a_{i2}^f) \nonumber \\[0.3cm]
 \Delta T_{20} & = & - \frac{m_f'}{m_f |\kappa_{A^0}^f| } \,
 \left\{ \begin{array}{rl}
 \mu - A_u \cot \beta  & , \ f = d \\
 \mu - A_d \tan \beta  & , \ f = u
 \end{array} \right\} \cdot  \nonumber \\ & &
 \cdot \, \sum_{i = 1}^4 \,
 V_{5}^p \, (k^2,m_f,m_{\tilde{f}_1},m_{\tilde{f}_2},m_{\tilde{\chi}_i^0},
 v_{i1}'^f+a_{i1}'^f,v_{i1}'^f-a_{i1}'^f,
 v_{i2}'^f-a_{i2}'^f,v_{i2}'^f+a_{i2}'^f)  \ , \nonumber
\end{eqnarray}
where
\begin{eqnarray*}
N'_{j1} = N_{j1} \, c_W + N_{j2} \, s_W \ , \ N'_{j2} = - N_{j1} \, s_W +
N_{j2} \, c_W  \ .
\end{eqnarray*}
The chargino and neutralino mass matrix $V_{ij}$, $U_{ij}$, $N_{ij}$
are given in appendix B. The vertex correction diagrams are
described by the following functions with masses and couplings in its
arguments.
\begin{eqnarray*}
 \lefteqn{
 V_1^{s,p} (k^2,m_f,m_1,m_2,m_3,v,a) \  = } \\ & & 4 \ [ \ ( \, \pm m_1^2 \
 (v^2 - a^2) \mp m_f m_1 \ ( v^2 + a^2 ) + ( m_f^2 - k^2/2 )
 ( v^2 - a^2 ) ) \ C_0 \\  \nonumber  & &
 + ( \  m_f m_1 (1 \pm 1) \  ( v^2 + a^2 ) - ( 4 m_f^2 - k^2 )
 ( v^2 - a^2 ) \ ) \ C_1^+ + ( v^2 - a^2 ) \\ \nonumber  & &
 ( 4 C_{20} + ( 4 m_f^2 - k^2 ) \ C_2^+
 + k^2 C_2^- - 1 ) \ ] \, (k^2,m_1,m_2,m_3)  \\[0.2cm]  \nonumber
 \lefteqn{
 V_2^{s,p} (k^2,m_f,m_1,m_2,m_3,v,a) \mbox{}  = } \\ & &  [ \  ( \  ( m_f^2 +
 m_1^2 )  ( v^2 - a^2 ) + 2 m_f m_1 \  ( v^2 + a^2 ) \  ) \  C_0 \mp
 4 m_f^2 ( v^2 - a^2 ) \ C_1^+ \\ \nonumber  & &
 - 2 \, (1 \pm 1) \, m_f m_1 \ ( v^2 + a^2 ) \ C_1^+ \pm ( v^2 - a^2 ) \
 ( 4 C_{20} + ( 4 m_f^2 - k^2 ) C_2^+ \\ \nonumber  & & + k^2 C_2^- - 1/2 )
   \ ] (k^2,m_1,m_2,m_3)  \\[0.2cm]  \nonumber
 \lefteqn{
 V_3^{s,p} (k^2,m_f,m_1,m_2,m_3,v,a,v',a',v'',a'') = } \\ & &
 - \, [ \  ( m_f^2 ( \pm v v' a'' + a a' v'' ) \  +  m_f m_2 \
 ( \pm v a ' a'' + a v' v'' ) \\ \nonumber  &  & +
 m_f m_1 \ ( v a' v'' \pm a v' a'' ) + m_1 m_2 \ ( v v' v'' \pm
 a a' a'' ) ) \ C_0\\ \nonumber & &    - 2 m_f ( m_2 \ ( \pm
 v a ' a'' + a v' v'') + m_1 \ ( v a' v'' + a v' a'' ) + \\ \nonumber  & &
 2 m_f \ ( \pm v v' a'' + a a' v'' ) ) \ C_1^+  +  ( v v' a'' \pm a a' v'' ) \
 ( 4 C_{20} + \\ \nonumber  & & ( 4 m_f^2 - k^2 ) \ C_2^+ + k^2 C_2^- - 1/2
 ) \ ] \ (k^2,m_1,m_2,m_3)  \\[0.2cm]   \nonumber
\lefteqn{
 V_4^s (k^2,m_f,m_1,m_2,m_3,v,a,v',a')  = } \\ & &  [ \ 2 \ m_f m_3 \ ( v v' +
 a a' )  \ ( C_0 - 2 C_1^+ ) + 4 \ ( v v' - a a' )
    \\ \nonumber &  &  ( \ (m_f^2 - k^2 ) \ C_1^+ -  k^2 C_1^- \ )
 - 2 \ ( v v' - a a' ) \ ( \ 4 C_{20} -
  \\ \nonumber &  &  1/2 + ( 4 m_f^2 - k^2  ) \  C_2^+
 + k^2 C_2^-  ) \  ] \ (k^2,m_1,m_2,m_3)  \\[0.2cm]  \nonumber
 \lefteqn{
 V_4^p (k^2,m_f,m_1,m_2,m_3,v,a,v',a')  = } \\ & &  [ \ 6 \ m_f m_3 \ ( v a'
 + a v' )  \ C_0  + 2 \ ( a v' - v a' ) \ ( \ 4 C_{20} -
  \\ \nonumber &  &  1/2 + ( 4 m_f^2 - k^2  ) \  C_2^+
 + k^2 C_2^-  ) \  ] \ (k^2,m_1,m_2,m_3)  \\[0.2cm]  \nonumber
 \lefteqn{
 V_5^s (k^2,m_f,m_1,m_2,m_3,v,a,v',a')  = } \\ & & - \, [ \ 2
 m_f \ ( v v' - a a' )
 \ C_1^+  + m_3 \  ( v v' + a a' ) \  C_0 \ ] \  (k^2,m_1,m_2,m_3)
 \\[0.2cm]  \nonumber
 \lefteqn{
 V_5^p (k^2,m_f,m_1,m_2,m_3,v,a,v',a') = } \\ & & - \, m_3
 ( v a' + a v' ) \, C_0 \ (k^2,m_1,m_2,m_3)  \\[0.2cm]  \nonumber
 \lefteqn{
 V_6^s (k^2,m_f,m_1,m_2,m_3,v,a,v',a')  = } \\ & & - 4 \ [ \ m_3 \ ( v v' -
 a a'
 )  C_0  -  m_{f} \ ( v v' + a a' ) \ C_1^+ \ ] \
 (k^2,m_1,m_2,m_3) \ . \nonumber \\[0.2cm] \nonumber
\end{eqnarray*}
%
%
\smallskip
Fermion self energies are represented by the scalar functions
$\Sigma^f_{S,V,A}$ through the decomposition in the scalar,
vector and axialvector part:
\begin{equation}
 \Sigma^f (k) = \not{k} \Sigma_V^f (k^2) + \not{k} \gamma_5
  \Sigma_A^f (k^2) + m_f \Sigma_S^f (k^2)  \, .
\end{equation}
Standard model and genuine SUSY self energy contributions are listed
separately in the following:
\begin{eqnarray}
\Sigma_{S,2-Higgs}^f (k^2)  & = & - \frac{\alpha }{4 \pi } \ [ \ (v_f^2 -
  a_f^2) \, (4 B_0 (k^2,m_f,M_Z) - 2)
 -\frac{m_{f'}^2}{2 s_W^2 M_W^2} \nonumber \\ & &
 ( \, B_0 (k^2,m_{f'},M_W) + B_0 (k^2,m_{f'},M_{H^+}) \,)
 -\frac{m_f^2}{4 s_W^2 M_W^2} \nonumber \\ & & ( ( \kappa_{h^0}^f )^2
 B_0 (k^2,m_f,M_{h^0}) + ( \kappa_{H^0}^f )^2 B_0 (k^2,m_f,M_{H^0})
 \nonumber \\ & & -
 | \kappa_{A^0}^f |^2 B_0 (k^2,m_f,M_{A^0}) -  B_0 (k^2,m_f,M_Z) \,)
 \ ] \nonumber \\[0.3cm]
\Sigma_{S, SUSY}^f (k^2) & = & \frac{\alpha }{4 \pi} \
 [ \ \sum_{i=1}^{2} \, \frac{m_{\tilde{\chi}^+_i}}{m_f s_W^2} \, (
 v_{i1}^f a_{i1}^f \, B_0 (k^2, m_{\tilde{\chi}_i^+}, m_{\tilde{f}'_1} )
 + v_{i2}^f a_{i2}^f \,
 B_0 (k^2, m_{\tilde{\chi}_i^+}, m_{\tilde{f}'_2}) \, )  \nonumber \\ & &
 + \, \sum_{i=1}^4 \, \frac{2 m_{\tilde{\chi}^0_i}}{m_f}
 ( \, v_{i1}'^f a_{i1}'^f \,
 B_0 (k^2, m_{\tilde{\chi}_i^0}, m_{\tilde{f}_1}) +
 v_{i2}'^f a_{i2}'^f \,
 B_0 (k^2, m_{\tilde{\chi}_i^0}, m_{\tilde{f}_2}) \, ) \ ] \nonumber \\
%
%
 \Sigma_{V,2-Higgs}^f (k^2) & = & - \frac{\alpha }{4 \pi } \ [ \ (v_f^2 +
 a_f^2) (2 B_1 (k^2,m_f,M_Z) + 1 ) + \frac{1}{4 s_W^2} \nonumber \\ & &
 (2 B_1(k^2,m_{f'},M_W) + 1)
 + \frac{m_f^2}{4 s_W^2 M_W^2}\, (\, (\kappa_{h^0}^f)^2 B_1(k^2,m_f,M_{h^0})
 \nonumber \\ & &
+ (\kappa_{H^0}^f)^2 B_1(k^2,m_f,M_{H^0}) +  B_1(k^2,m_f,M_Z) +
 | \kappa_{A^0}^f |^2 \nonumber \\ & & B_1(k^2,m_f,M_{A^0}) \, )
 + \frac{1}{4 s_W^2 M_W^2} \ ( \ (m_f^2 + m_{f'}^2) B_1 (k^2,m_{f'},M_W)
 \nonumber \\ & &
 +  \frac{m_f^2 \tan^2  \beta + m_{f'}^2 \cot^2 \beta}{4 s_W^2 M_W^2} \,
 B_1 (k^2,m_{f'},m_{H^+}) \, ) \ ]  \nonumber \\
 \nonumber \\[0.2cm]
 \Sigma_{V, SUSY}^f (k^2) & = & - \frac{\alpha }{4 \pi } \,
 \frac{1}{s_W^2} \ [ \ \sum_{i=1}^{2} \,
 \frac{ (v_{i1}^f)^2 + (a_{i1}^f)^2 }{2} \,
 B_1(k^2, m_{\tilde{\chi}_i^+}, m_{\tilde{f}'_1}) +
 \frac{ (v_{i2}^f)^2 + (a_{i2}^f)^2 }{2} \nonumber \\ & &
 B_1(k^2, m_{\tilde{\chi}_i^+}, m_{\tilde{f}'_2})
 + 2 s_W^2 \sum_{i=1}^{4} \, ( \,
  \frac{ (v_{i1}'^f)^2 + (a_{i1}'^f)^2 }{2} \, \nonumber \\ & &
 B_1(k^2, m_{\tilde{\chi}_i^+}, m_{\tilde{f}_1}) +
 \frac{ (v_{i2}'^f)^2 + (a_{i2}'^f)^2 }{2} \,
 B_1(k^2, m_{\tilde{\chi}_i^+}, m_{\tilde{f}_2}) \, ) \ ] \nonumber
\end{eqnarray}
\begin{eqnarray}
\Sigma_{A,2-Higgs}^f (k^2) & = & - \frac{\alpha }{4 \pi } \ [ \ - 2 v_f a_f
 (2 B_1 (k^2,m_f,M_Z) + 1 ) - \frac{1}{4 s_W^2}  \nonumber \\ & &
 (2 B_1(k^2,m_{f'},M_W) - 1)  + \frac{ 1 }{4 s_W^2 M_W^2} \, ( \,
 (m_{f'}^2 - m_f^2) B_1 (k^2,m_{f'},M_W) \nonumber \\ & &
 + (m_f^2 \tan^2 \beta - m_{f'}^2 \cot^2 \beta ) B_1 (k^2,m_{f'},m_{H^+}) \, )
 \ ] \nonumber \\[0.2cm]
 \Sigma_{A, SUSY}^f (k^2) & = & - \frac{\alpha }{4 \pi } \, \frac{1}{s_W^2}
 \ [ \ \sum_{i=1}^2 \, (
 \frac{ -(v_{i1}^f)^2 + (a_{i1}^f)^2 }{2} \,
 B_1(k^2, m_{\tilde{\chi}_i^+}, m_{\tilde{f}'_1}) +
 \frac{ -(v_{i2}^f)^2 + (a_{i2}^f)^2 }{2} \nonumber \\ & &
 B_1(k^2, m_{\tilde{\chi}_i^+}, m_{\tilde{f}'_2}) \,)
 + 2 s_W^2 \, \sum_{i=1}^4 \, ( \,
  \frac{ -(v_{i1}'^f)^2 + (a_{i1}'^f)^2 }{2} \,
 B_1(k^2, m_{\tilde{\chi}_i^0}, m_{\tilde{f}_1})\nonumber \\ & &
 + \frac{-(v_{i2}'^f)^2 + (a_{i2}'^f)^2 }{2}
 B_1(k^2, m_{\tilde{\chi}_i^0}, m_{\tilde{f}_2}) \, ) \ ] \ .
\end{eqnarray}
The 2-point functions $B_0$, $B_1$ are defined
\begin{eqnarray}
 B_0(k^2,m_1,m_2) & = & \Delta - \int_0^1 \, dx \,\log\frac
    {x^2k^2 -x(k^2+m_1^2-m_2^2)+m_1^2-i\epsilon}{\mu^2}  \ , \nonumber \\
 B_1(k^2,m_1,m_2) & = & - \frac{k^2+m_1^2-m_2^2}{2k^2}\,B_0(k^2,m_1,m_2)
       +  \frac{m_1^2-m_2^2}{2k^2}\, B_0(0,m_1,m_2) \ , \nonumber \\
\end{eqnarray}
where
$$ \Delta = \frac{2}{\varepsilon} - \gamma + \log 4\pi, \;\;\;\;\;\;
    \varepsilon = 4-D \, ,  $$
and the mass scale $\mu$ are the UV-parameters from dimensional
regularization in D-dimensions.
\smallskip \par
The 3-point functions $C_0$, $C_1^{+,-}$, $C_2^{0,+,-}$ are
for equal external fermion masses $ p^2 = p^{'2} = m_f^2 $ :
\begin{eqnarray}
\frac{i}{(4 \pi )^2}\ C_0 \ (k^2,m_1, m_2, m_3) & = &
             \int \ \frac{d^4 k}{(2 \pi)^4}
 \frac{1}{D_1 D_2 D_3} \ ,
\end{eqnarray}
with $ k^2 = (p - p')^2 $ and the denominators
\begin{eqnarray}
D_1 & = & (k - p')^2 - m_1^2 + i \epsilon \nonumber \\
D_2 & = & (k - p)^2 - m_2^2 + i \epsilon \nonumber \\
D_3 & = & k^2 - m_3^2 + i \epsilon \, . \nonumber
\end{eqnarray}
and for different masses $ m_1 , \ m_2 , \ m_3 $ :
\begin{eqnarray}
(p + p')^2  C_1^+ (k^2,m_1,m_2,m_3) & = &
 B_0 (k^2, m_1, m_2) - \frac{B_0(m_f^2, m_1, m_3) + B_0(m_f^2, m_2, m_3)}
 {2} + \nonumber \\ \nonumber  & & + \frac{
 2 m_f^2 + 2 m_3^2 - m_1^2 - m_2^2}{2}  \ C_0  \nonumber \\[0.2cm]  \nonumber
 2 k^2 C_1^- (k^2,m_1,m_2,m_3)  & = & B_0(m_f^2, m_1, m_3) -  B_0(m_f^2, m_2,
 m_3) + ( m_2^2 - m_1^2 ) \ C_0 \\[0.2cm] \nonumber
 4 C_2^0 (k^2,m_1,m_2,m_3) & = & B_0 (k^2, m_1, m_2) + ( m_1^2 + m_2^2 - 2
 m_3^2 - 2 m_f^2 ) \ C_1^+ \\ \nonumber  & & + (m_1^2 - m_2^2) C_1^- +  2
 m_3^2 C_0 + 1  \\[0.2cm] \nonumber
(p + p') ^2 C_2^+ (k^2,m_1,m_2,m_3) & = & \frac{1}{4}
 [B_1 (m_f^2, m_3, m_2) + B_1 (m_f^2, m_3, m_1) + 2 B_0 (k^2, m_1, m_2)
 \\ \nonumber & & + 2 (2 m_3^2 - m_1^2 - m_2^2 + 2 m_f^2) C_1^+ ]
 - C_2^0  \\[0.2cm] \nonumber
 k^2 C_2^- (k^2,m_1,m_2,m_3)  & = & - \frac{1}{4} (
 B_1 (m_f^2, m_3, m_2) + B_1 (m_f^2, m_3, m_1) ) \\ \nonumber & &  + 2 \ (
 m_1^2 - m_2^2 )  C_1^-   - C_2^0 \ .
\end{eqnarray}
The analytic expression for the scalar vertex integral $C_0$ can be found
in \cite{bhs}.
\renewcommand{\thesubsection}{B}
\renewcommand{\theequation}{B.\arabic{equation}}
\subsection{Gaugino mass matrix}
\setcounter{equation}{0}\setcounter{footnote}{0}
\vspace*{0.1cm} \hspace*{0.4cm}
The chargino $2 \times 2$ mass matrix is given by
\begin{equation}
  \cal{M}_{\rm \tilde{\chi}^\pm} \rm = \left( \begin{array}{ll}
    M & M_W \sqrt{2} \sin \beta \\ M_W \sqrt{2} \cos \beta & - \mu \\
    \end{array}  \right) \ ,
\end{equation} \\
with the SUSY soft breaking parameters $\mu$ and $M$ in the diagonal
matrix elements. The physical chargino mass states $\tilde{\chi}^{\pm}_i$
are the rotated
wino and charged Higgsino states:
\begin{eqnarray}
\tilde{\chi}^+_i & = & V_{ij} \psi^+_j   \nonumber \\
\tilde{\chi}^-_i & = & U_{ij} \psi^-_j  \ ; \ i,j = 1,2  \ .
\end{eqnarray}
$V_{ij}$ and $U_{ij}$ are unitary chargino mixing matrices obtained from
the diagonalization of the mass matrix (B.1):
\begin{equation}
\rm U^* \cal{M}_{\rm \tilde{\chi}^\pm} \rm  V^{-1} =
diag(m_{\tilde{\chi}^\pm_1}^2,m_{\tilde{\chi}^\pm_2}^2) \ .
\end{equation}
\smallskip \par
The neutralino $4 \times 4 $ mass matrix yields:
\begin{equation}
 \cal{M}_{\rm \tilde{\chi}^0}   = \left( \begin{array}{cccc}
 M' & 0 & - M_Z \sin \theta_W \cos \beta & M_Z \sin \theta_W \sin \beta \\
 0 & M & M_Z \cos \theta_W \cos \beta & - M_Z \cos \theta_W \sin \beta \\
- M_Z \sin \theta_W \cos \beta & M_Z \cos \theta_W \cos \beta & 0 &  \mu \\
 M_Z \sin \theta_W \sin \beta & - M_Z \cos \theta_W \sin \beta & \mu & 0
 \\  \end{array} \right)
\end{equation}
where the diagonalization introduces the unitary matrix $N_{ij}$ by:
\begin{equation}
  \rm N^* \cal{M}_{\rm \tilde{\chi}^0} \rm N^{-1}  = diag(
  m_{\tilde{\chi}_i^0}) \ .
\end{equation}
%
\newpage
\newpage
{\bf FIGURE CAPTIONS}
\vskip 0.5cm
\noindent {\bf Figure 1.}~
The mixing angle $\sin^2 \alpha$ as a function of the physical light
Higgs mass $M_{h^0}$ for $\tan \beta = 2$, $30$ in Fig. 1a) and
$\tan \beta = 0.5$, $8$ in Fig. 1b). Figs. 1 c,d) show the same data for
$\sin^2 \alpha$ as a function of the pseudoscalar mass $M_{A^0}$.
The solid line is $\sin^2 \alpha_{eff}$ in the effective potential
approximation of Eq. (\ref{glalpheffx}).
$\sin^2 \alpha_{1-loop}$ in the full one-loop
calculation of Eq. (\ref{glsina1}) is shown by the dotted line and
$\sin^2 \alpha_{b}$ in Eq. (\ref{alphaf}) with the
non-universal $h^0 \rightarrow b \bar{b}$ vertex correction is the
dashed result. The parameters in Fig. 1 are $m_t = 175$ GeV,
$m_{sf} = 700$ GeV, $\mu = 100$ GeV, $M = 550$ GeV, no sfermion left-right
mixing.
\vskip 0.5cm
\noindent {\bf Figure 2.}~
Figs. 2a,b) show the decay width $\Gamma_{h^0 \rightarrow b  \bar{b}}$,
including the full weak MSSM one-loop corrections
as a function of the physical light MSSM Higgs mass $M_{h^0}$ for
$\tan \beta = 2$, $30$ in Fig. 2a) and $\tan \beta = 0.5$, $8$ in
Fig. 2b). $m_t = 175$ GeV (solid line), $m_t = 160$ GeV (dotted line) and
$m_t = 190$ GeV (dashed line). No sfermion mixing.
The standard Higgs decay width
$\Gamma_{h^0 \rightarrow b  \bar{b}}$ with one-loop electroweak corrections
is labeled in Fig. 2.
Figs. 2c,d) contain the data of Figs. 2a,b) as a function of the
pseudoscalar Higgs mass $M_{A^0}$. The standard Higgs mass is
chosen to be equal the physical light Higgs mass $M_{h^0}$.
\vskip 0.5cm
\noindent {\bf Figure 3.}~
The $h^0 \rightarrow b \bar{b}$ decay width dependence on the
sfermion masses is plotted in Figs. 3a,b) as a function of
the physical light MSSM Higgs mass $M_{h^0}$.
$\tan \beta = 2$, $30$ in Fig. 3a) and
$\tan \beta = 0.5$, $8$ in Fig. 3b). All sfermion soft breaking
parameters are equal, $m_{sf} = 1$ TeV (solid line),
$m_{sf} = 500$ GeV (dotted line), $m_{sf} = 300$ GeV (short dashed line)
and  $m_{sf} = 200$ GeV (long dashed line) and no sfermion mixing is
assumed.
In Fig. 3c) mixing effects of sfermions are shown for $\tan \beta =
2$, $30$. $A_t' = 0$ (solid line), $A_t' = 100$ GeV (dotted line),
$A_t' = 200$ GeV (short dashed), $A_t' = 300$ GeV (long dashed) and
$A_t' = 400$ GeV (dot dashed), see Eq. (\ref{glaprime}).
Gaugino contributions to the decay width $h^0 \rightarrow b \bar{b}$
are plotted in Fig. 3d) for $M = 100$ GeV (dotted) and
$M = 200$ GeV (dashed). The $\mu$ parameters are described in the
figure.
\vskip 0.5cm

\vskip 0.5cm
\noindent {\bf Figure 4.}~
Figs. 4 a,b) show
the $H^0 \rightarrow b \bar{b}$ decay width with full MSSM one-loop
contributions for $\tan \beta = 0.5$, $2$, $8$, $30$ as a function
of the physical heavy Higgs mass $M_{H^0}$. In Fig. 4a) the top-quark
mass is $ m_t = 175$ GeV (solid line), $m_t = 160 $ GeV
(dotted line) and $m_t = 190$ GeV (dashed line). The soft breaking
parameters are $m_{sf} = 700$ GeV, $\mu = 100$ GeV, $M = 550$ GeV.
Fig. 4b) shows the dependence on sfermion masses for a constant
top-quark mass $m_t = 175$ GeV.  The lines correspond to
$m_{sf} = 1$ TeV (solid), $m_{sf} = 500$ GeV (dotted),
$m_{sf} = 300$ GeV (short dashed), $m_{sf} = 200$ GeV (long dashed).
The pseudoscalar $A^0 \rightarrow b \bar{b}$ decay width is plotted
in Fig. 4c) as a function of the pseudoscalar mass $M_{A^0}$ with full
MSSM one-loop contributions. Fig. 4d) shows
the $A^0 \rightarrow b \bar{b}$ vertex corrections
$\delta \Gamma_b = 2 \Re e \Delta T_{A^0}$, Eq. (\ref{gamma1}),
for $\tan \beta = 0.5$, $2$, $30$ and top-quark masses $m_t =
175$ GeV (solid line),
$m_t = 160$ GeV (dotted line), $m_t = 190$ GeV (dashed line). In Figs.
4 c,d) $m_{sf} = 700$ GeV, $\mu = 100$ GeV, $M = 550$ GeV.
\vskip 0.5cm
\noindent {\bf Figure 5.}~
Gluino contributions to the one-loop vertex corrections are plotted
in Figs. 5 a,b) for the $h^0 \rightarrow b \bar{b}$ decay width,
Figs. 5 c,d ) for $H^0 \rightarrow b \bar{b}$ and in Figs. 5 e,f)
for $A^0 \rightarrow b \bar{b}$. In Figs. 5 a,c,e) the
$H \rightarrow b \bar{b}$ decay width is plotted as a function of the
respective Higgs mass. $\tan \beta = 0.5$, $2$, $8$, $30$ with a
gluino mass $m_{gl} = 500$ GeV (solid line), $m_{gl} = 200$ GeV (dotted line).
Sfermion soft breaking masses are  $m_{sf} = 700$ GeV (solid line) and
$m_{sf} = 500$ GeV (dotted line),
$m_t = 175$ GeV, $\mu = 100$ GeV, $M = 550$ GeV.
Figs. 5 b,d,f) show the $\mu$ parameter dependence on the
$H \rightarrow b \bar{b}$ vertex corrections for pseudoscalar
masses $M_{A^0} = 50$ GeV, $150$ GeV, $250$ GeV. $\tan \beta = 8$,
$m_t = 175$ GeV, $m_{gl} = 200$ GeV, $m_{sf} = 500$ GeV, $M = 550$ GeV
and $A_f = 0$.
\vskip 0.5cm
\noindent {\bf Figure 6.}~
The light Higgs boson decay branching ratios $h^0 \rightarrow f \bar{f}$
for the decay channels $f = b$ (Fig. 6 a), $\tau$, $c$ (Fig. 6 b)
with full one-loop MSSM contributions (solid line) and in the
approximation of Eq. (\ref{glalpheffx}) (dashed).
The branching ratios are plotted as functions of the physical light
Higgs mass $M_{h^0}$ for $\tan \beta = 0.5$, $2$, $8$, $30$.
The dotted curves are the standard Higgs branching ratios.
$m_t = 175$ GeV, $m_{sf} = 700$ GeV, $\mu = 100$ GeV, $M = 550$ GeV,
no sfermion left-right mixing.
\newpage
\begin{figure}
\epsfig{figure=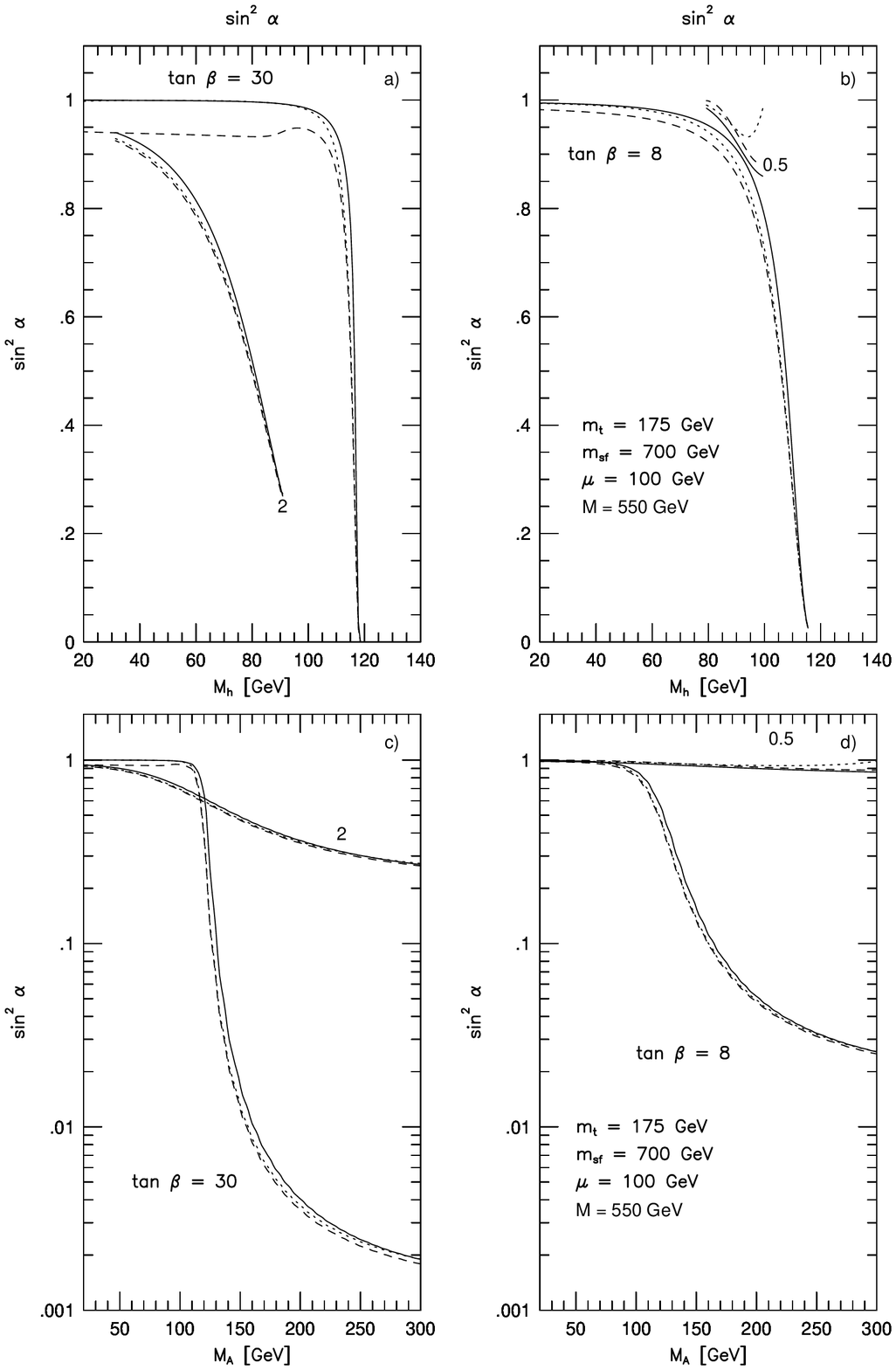,%
height=19cm,width=18.0cm,%
bbllx=75pt,bblly=75pt,bburx=555pt,bbury=751pt}
\caption{}
\end{figure}
\newpage
\begin{figure}
\epsfig{figure=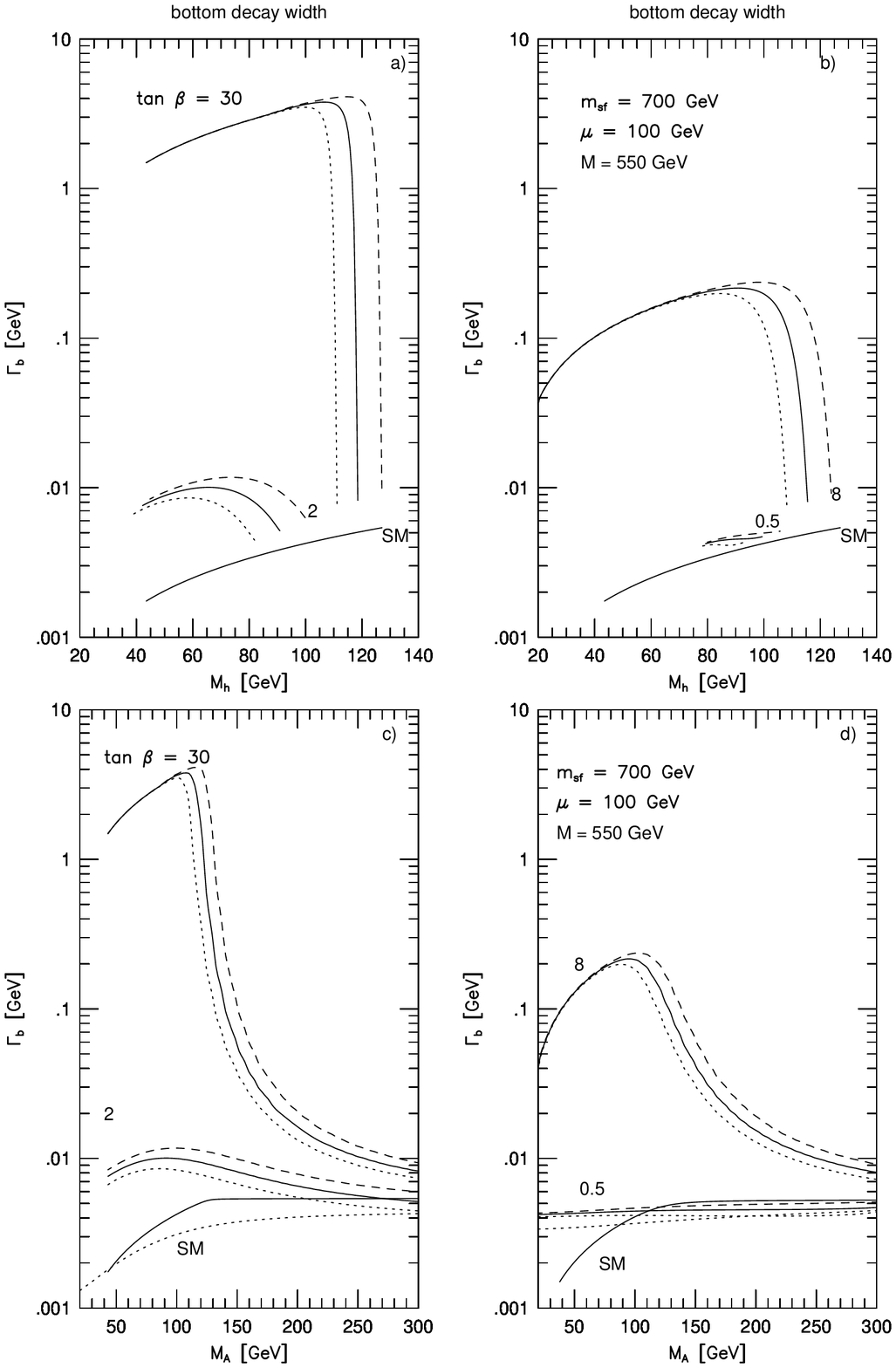,%
height=19cm,width=18.0cm,%
bbllx=75pt,bblly=75pt,bburx=555pt,bbury=751pt}
\caption{}
\end{figure}
\newpage
\begin{figure}
\epsfig{figure=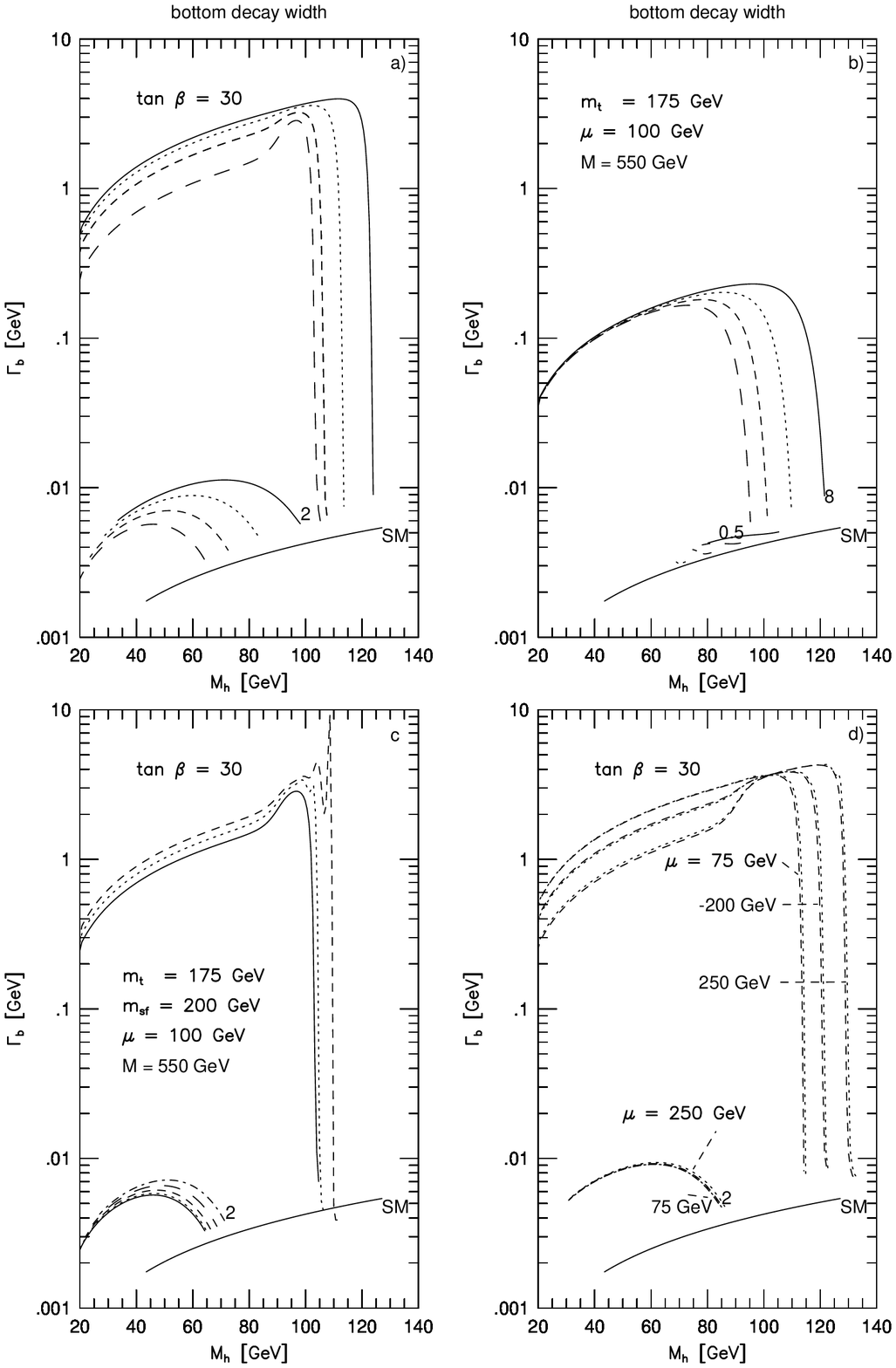,%
height=19cm,width=18.0cm,%
bbllx=75pt,bblly=75pt,bburx=555pt,bbury=751pt}
\caption{}
\end{figure}
\newpage
\begin{figure}
\epsfig{figure=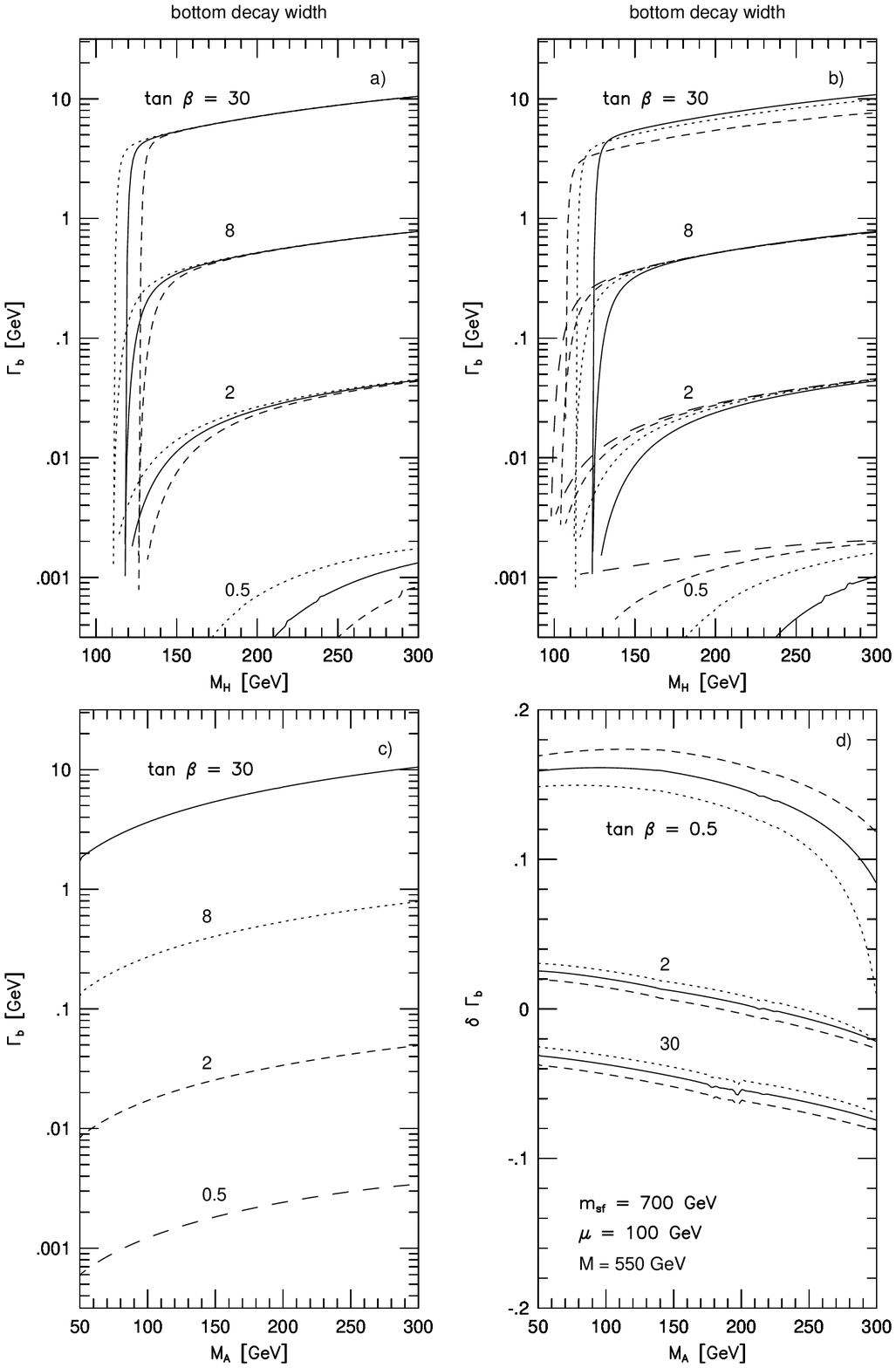,%
height=19cm,width=18.0cm,%
bbllx=75pt,bblly=75pt,bburx=555pt,bbury=751pt}
\caption{}
\end{figure}
\newpage
\begin{figure}
\epsfig{figure=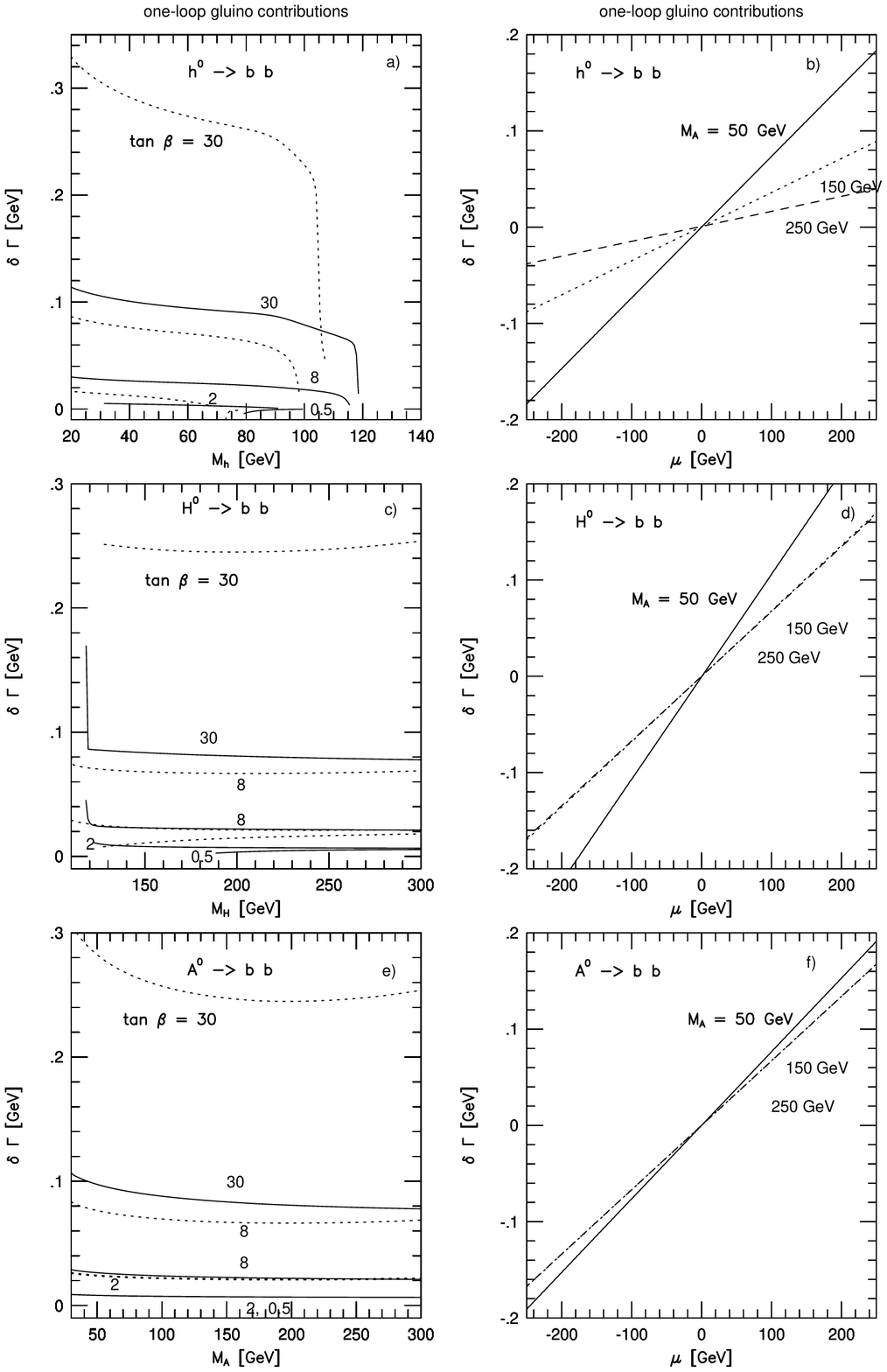,%
height=19cm,width=18.0cm,%
bbllx=75pt,bblly=75pt,bburx=555pt,bbury=751pt}
\caption{}
\end{figure}
\newpage
\begin{figure}
\epsfig{figure=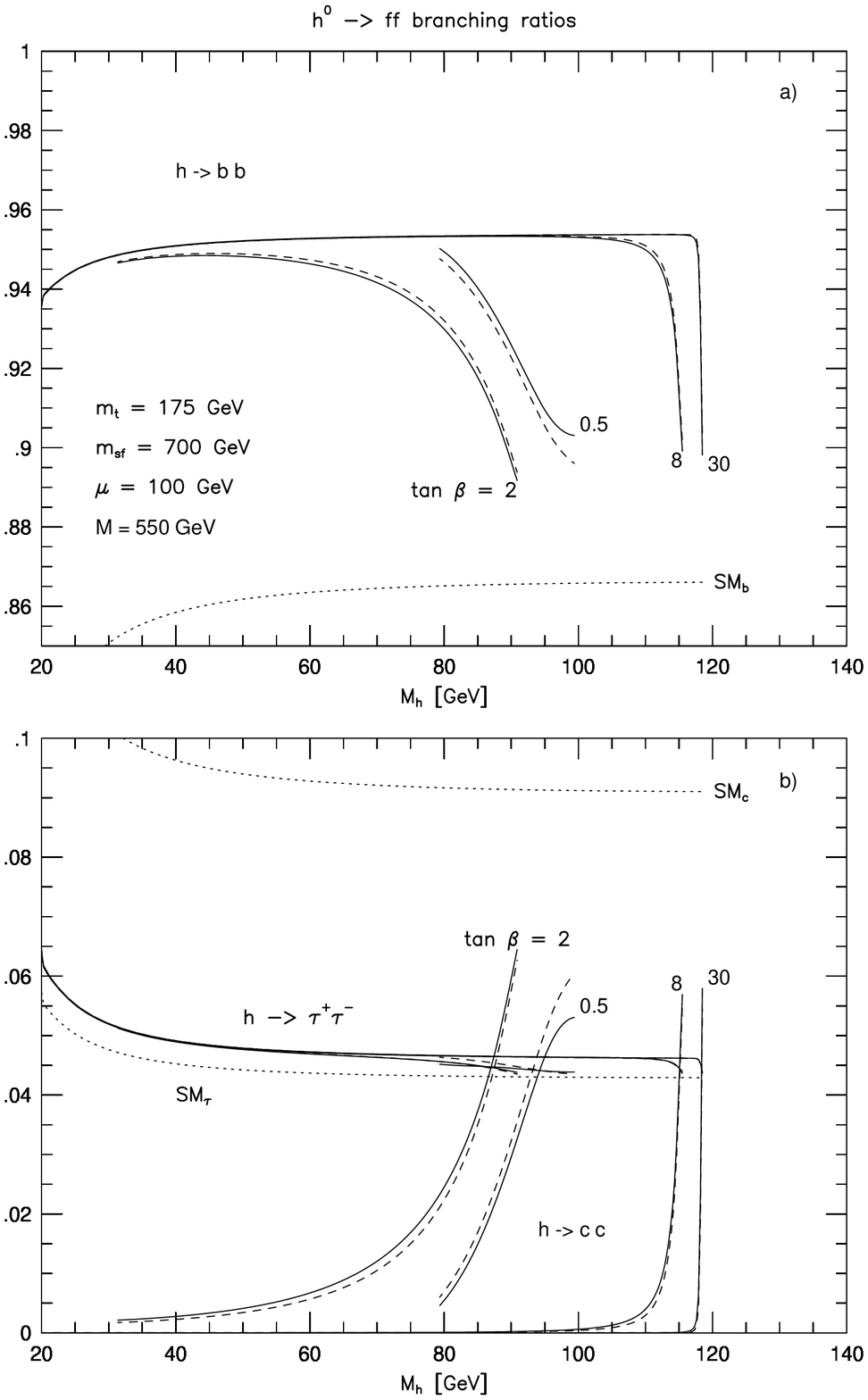,%
height=19cm,width=18.0cm,%
bbllx=75pt,bblly=75pt,bburx=555pt,bbury=751pt}
\caption{}
\end{figure}

\begin{thebibliography}{xxx}
%
\bibitem{hunter}  H.P. Nilles, Phys. Rep. \bf 110  \rm (1984) 1\\
      H.E. Haber, G. Kane , Phys. Rep. \bf 117  \rm (1985) 75\\
      J.F. Gunion, H.E. Haber, Nucl. Phys. \bf B 272  \rm (1986) 1;
      Nucl. Phys. \bf B 402 \rm (1993) 567\\
      J.F. Gunion, H.E. Haber, G. Kane, S. Dawson:
      The Higgs Hunter's Guide,  Addison-Wesley 1990
\bibitem{boer}  J. Ellis, S. Kelly, D. Nanopoulos, Phys. Lett. \bf B 249
                \rm (1990) 441 \\
                U. Amaldi, W.de. Boer, H. F\"urstenau, Phys. Lett. \bf
                B 260 \rm (1991) 447 \\
                P. Langacker, M. Luo, Phys. Rev. \bf D 44 \rm (1991) 477
\bibitem{hara} L. Hall, L. Randall, Phys. Rev. Lett. \bf 65 \rm
               (1990) 2939
\bibitem{hempf1} H.E. Haber, R. Hempfling, Phys. Rev. Lett. \bf 66  \rm
                 (1991) 1815
\bibitem{upper1} Y. Okada, M. Yamaguchi, T. Yanagida, Prog. Theor. Phys.
                 \bf 85 \rm (1991) 1 \\
                 D. Pierce, A. Papadopoulos, S. Johnson, Phys. Rev. Lett.
                 \bf 68 \rm (1992) 3678
\bibitem{barb1} R. Barbieri, M. Frigeni, Phys. Lett. \bf 258 B \rm
                (1991) 395 \\
                R. Barbieri, M. Frigeni, F. Caravaglios, Phys. Lett.
                \bf 258 B \rm (1991) 167 \\
                J.R. Espinosa, M.Quiros, Phys. Lett. \bf 266 \rm
                (1991) 389 \\
                K. Sasaki, M. Carena, C.E.M. Wagner, Phys. Rev. Lett. \bf
                381 B \rm (1992) 66\\
                H.E. Haber, R. Hempfling, Phys. Rev. \bf D 48 \rm
                (1993) 4280
\bibitem{ellis} J. Ellis, G. Ridolfi, F. Zwirner, Phys. Lett. \bf 257 B \rm
                (1991) 83 \\
                J. Ellis, G. Ridolfi, F. Zwirner, Phys. Lett. \bf 262 B \rm
                (1991) 477
\bibitem{okada} Y. Okada, M. Yamaguchi, T. Yanagida, Phys. Lett. \bf 262 B
                \rm (1991) 54 \\
                J.L. Lopez, D.V. Nanopoulos, Phys. Lett. \bf 266 B \rm (1991)
                397 \\
                A. Brignole, J. Ellis, G. Ridolfi, F. Zwirner,
                Phys. Lett. \bf 271 B \rm (1991) 123
\bibitem{yama}  A. Yamada, Phys. Lett. \bf 263 B \rm (1991) 233 \\
                M.A. Diaz, H.E. Haber, Phys. Rev. \bf D 45 \rm (1992) 4246 \\
                M.A. Diaz, H.E. Haber, Phys. Rev. \bf D 46 \rm (1992) 3086 \\
                A. Brignole, Phys. Lett. \bf 281 B \rm (1992) 284 \\
                A. Brignole, Phys. Lett. \bf 277 B \rm (1992) 313 \\
                D. Pierce, A. Papadopoulos, Phys. Rev. \bf D 47 \rm
                (1992) 222 \\
                A. Yamada, Z. Phys. \bf C 61 \rm (1994) 247
\bibitem{upper2} J. Kodaira, Y. Yasui, K. Sasaki, Hiroshima preprint
                 HUPD-9316, YNU-HEPTh-93-102 (Nov. 1993) \\
                 R. Hempfling, A. H. Hoang, Phys. Lett. \bf B 331 \rm
                 (1994) 99  \\
                 J.A. Casas, J.R. Espinosa, M. Quiros, A. Riotto,
                 CERN-TH. 7334/94 (July 1994)
\bibitem{pok1} P.H. Chankowski, S. Pokorski, J. Rosiek, Phys. Lett.
               \bf 274 B \rm (1992) 191 \\
               P.H. Chankowski, S. Pokorski, J. Rosiek,
               Nucl. Phys. \bf 423 B \rm (1994) 437
\bibitem{dabx} A. Dabelstein, Karlsruhe preprint KA-THEP-5-1994,
               to appear in Z. Phys. \bf C \rm
\bibitem{pok6} P.H. Chankowski, S. Pokorski, J. Rosiek, Nucl. Phys. \bf
               423 B \rm (1994) 497
\bibitem{braaten} E. Braaten, J.P. Leveille, Phys. Rev. \bf D 22 \rm (1980)
                  715 \\
                  N. Sakai, Phys. Rev. \bf D 22 \rm (1980) 2220 \\
                  T. Inami, T. Kubota, Nucl. Phys. \bf B 179 \rm (1981) 171
\bibitem{bardin} D.Yu.\ Bardin, B.M. Vilensky, P.Ch. Christova,
                 Yad.\ Fiz.\ \bf 53\rm \ (1991) 240 (in russian)
\bibitem{Drees2} M. Drees, K. Hikasa, Phys. Lett. \bf 240 B \rm (1990) 455
\bibitem{spira1} A. Djouadi, M. Spira, P.M. Zerwas,
                 Phys. Lett. \bf B 276 \rm (1992) 350
\bibitem{spira2} A. Djouadi, M. Spira, J.J. van der Bij, P.M. Zerwas,
                 Phys. Lett. \bf B 257 \rm (1991) 187
\bibitem{haber2} J.Gunion, H.Haber, Nucl. Phys. \bf B 307 \rm (1988)
                 445
\bibitem{gluinos} K. Ng, H. Pois, T.C. Yuan, Phys. Rev. \bf D 40 \rm
                  (1989) 1689 \\
                  A. Djouadi, M. Drees, Madison-preprint MADPH-94-853,
                  October 1994
\bibitem{sirlin} A. Sirlin, Phys. Rev. \bf D 22 \rm (1980) 971  \\
                 W.J. Marciano, A. Sirlin, Phys. Rev. \bf D 22 \rm (1980)
                 2695
\bibitem{sola} D. Garcia, J. Sol\`{a}, Mod. Phys. Lett. \bf A 9 \rm
               (1994) 211 \\
               P.H. Chankowski, A. Dabelstein, W. Hollik, W. M\"osle,
               S. Pokorski, J. Rosiek, Nucl. Phys. \bf B 417 \rm
               (1994) 101
\bibitem{kunszt} Z. Kunszt, F. Zwirner, Nucl. Phys. \bf B 385 \rm
                 (1992) 3
\bibitem{diaz2}  M.A. Diaz, Vanderbilt-preprint VAND-TH-94-19, August 1994,
                 presented at the Eight Meeting of the Division of Particles
                 and Fields of the American Physical Society DPF '94, The
                 University of New Mexico, Albuquerque NM, August 2-6, 1994
\bibitem{jones}  D.R.T. Jones, Phys. Rev. \bf D 25 \rm (1984) 581 \\
                 O.V. Tarasov, A.A. Vladimirov, A.Y. Zharkov,
                 Phys. Lett \bf 93 B \rm (1980) 429 \\
                 S.G. Gorishny, A.L. Kataev, S. Larin, Sov. J. Nucl. Phys.
                 \bf 40 \rm (1984) 329 \\
                 D.V. Nanopoulos, D.A. Ross, Nucl. Phys. \bf B 157 \rm
                 (1979) 273 \\
                 R. Tarrach, Nucl. Phys. \bf B 183 \rm (1981) 384
\bibitem{top} F. Abe et al., CDF Collaboration, FERMILAB-PUB-95-022-E,
              March 1995  \\
              S. Abachi et al., D0 Collaboration, FERMILAB-PUB-95-028-E,
              March 1995
\bibitem{dabh}  A. Dabelstein, W. Hollik, Z. Phys. \bf C 53 \rm (1992) 507 \\
                B. Kniehl, Nucl. Phys. \bf B 376 \rm (1992) 3
\bibitem{bhs}   W. Hollik, Fortschr. Phys. \bf 38 \rm (1990) 165
%
\end{thebibliography}
\end{document}